\documentclass[aps,prl,twocolumn,final,letterpaper]{revtex4}
\usepackage{graphicx}   
\usepackage{import}                         
\usepackage{amsmath} 
\usepackage{bm}
\usepackage{amssymb}
\usepackage{xspace}
\usepackage{layouts}
\usepackage{transparent}
\usepackage{multirow}
\usepackage{mdframed}
\usepackage{color}
\usepackage{xcolor,soul}
\usepackage{bm}
\usepackage{dsfont}
\usepackage{slashed}
\usepackage{braket}
\usepackage{svg}
\usepackage[export]{adjustbox}
\usepackage{hyperref}
\hypersetup{colorlinks,
	linkcolor={blue!75!black!80!yellow},
	citecolor={blue!75!black!80!yellow},
	urlcolor={blue!75!black!80!yellow}
}

\usepackage[capitalize,nameinlink]{cleveref}
\crefname{subequations}{Eqs.}{Eqs.} 
\Crefname{subequations}{Eqs.}{Eqs.}
\crefformat{subequations}{#2Eqs.~(#1)#3}
\Crefformat{subequations}{#2Eqs.~(#1)#3}
\crefname{page}{p.}{p.} 
\usepackage{natbib}

\usepackage{textcomp} 

\renewcommand{\paragraph}[1]{\vskip 1ex\noindent\textbf{#1.}~}
\usepackage[utf8]{inputenc}

\begin{document}
\rmfamily

\title{
Highly-confined and tunable plasmonics based on two-dimensional solid-state defect lattices
}

\author{Ali Ghorashi$^{1}$}
\author{Nicholas Rivera$^{1,4}$}
\author{Bowen Shi$^{1,6}$}
\author{Ravishankar Sundararaman$^{2}$}
\author{Efthimios Kaxiras$^{3, 4}$}
\author{John Joannopoulos$^{1, 5}$}
\author{Marin Soljačić$^{1}$}

\affiliation{$^{1}$ Department of Physics, Massachusetts Institute of Technology, Cambridge, MA 02139, USA\looseness=-1}
\affiliation{$^{2}$ Rensselaer Polytechnic Institute, 110 8th Street, Troy, New York 12180, USA\looseness=-1}
\affiliation{$^{3}$ School of Engineering and Applied Sciences, Harvard University, Cambridge, MA  02134, USA
\looseness=-1}
\affiliation{$^{4}$ Department of Physics, Harvard University, Cambridge, MA 02134, USA}
\affiliation{$^{5}$ Institute for Soldier Nanotechnologies, Massachusetts Institute of Technology, Cambridge, MA 02139, USA}
\affiliation{$^{6}$ State Key Laboratory for Mesoscopic Physics and Department of Physics, Peking University, Beijing 100871, P. R. China}

\date{April 2023}

\begin{abstract}
Plasmons, collective excitations of electrons in solids, are associated with strongly confined electromagnetic fields, with wavelengths far below the wavelength of photons in free space. This strong confinement promises the realization of optoelectronic devices that could bridge the size difference between photonic and electronic devices. However, despite decades of research in plasmonics, many applications remain limited by plasmonic losses, thus motivating a search for new engineered plasmonic materials with lower losses. A promising pathway for low-loss plasmonic materials is the engineering of materials with flat and energetically isolated metallic bands, which can strongly limit phonon-assisted optical losses, a major contributor to short plasmonic lifetimes. Such electronic band structures may be created by judiciously introducing an ordered lattice of defects in an insulating host material. Here, we explore this approach, presenting several low-loss, highly-confined, and tunable plasmonic materials based on arrays of carbon substitutions in hexagonal boron nitride (hBN) monolayers. From our first-principles calculations based on density functional theory (DFT), we find plasmonic structures with mid-infrared plasmons featuring very high confinements ($\lambda_{\text{vacuum}}/\lambda_{\text{plasmon}}$ exceeding 2000) and quality factors in excess of 1000. We provide a systematic explanation of how crystal structure, electronic bandwidth, and many-body effects affect the plasmonic dispersions and losses of these materials. The results are thus of relevance to low-loss plasmon engineering in other flat band systems.


\end{abstract}

\maketitle

%
The optoelectronic properties of two-dimensional (2D) materials are of major interest due to the qualitatively different physics of electron-photon interactions in reduced dimensions.  Accordingly, the search for stable 2D materials with specific optoelectronic properties has been a topic of intense research in recent years \cite{huang2022carbon, lonvcaric2018strong, gjerding2021recent}. In particular, the expansion of the repertoire of stable 2D materials has made strides in three directions:\\
(i) The tuning of geometric properties of van der Waals (vdW) bilayers and trilayers, most notably by creating moir\'{e} patterns \cite{lewandowski2019intrinsically};\\
(ii) the advent or prediction of 2D analogues of naturally occurring three dimensional metals \cite{sundararaman2020plasmonics, novoselov2005two};  and\\
(iii) the introduction of defects in common 2D materials \cite{huang2022carbon} to induce desired optoelectronic properties.\\

A major impetus 
has been to enable collective excitations,
such as plasmons \cite{boriskina2017losses}, phonon-polaritons \cite{dai2019phonon, rivera2019phonon} and exciton-polaritons \cite{novko2021ab},
with tailored dispersions, high confinements, and/or low losses \cite{khurgin2015ultimate}, which would enable a wide range of new light-matter interaction effects \cite{rivera2016shrinking, rivera2020light}. In the case of plasmons specifically, whose electromagnetic fields can be confined far below the free-space wavelength of photons, many of the promising applications envisioned for the field decades ago are still hindered to this day by loss. Strongly-confined and low-loss plasmonic excitations could lead to major advances for most envisioned applications of plasmonics in fields spanning photovoltaics \cite{atwater2010plasmonics}, spectroscopy \cite{langer2019present}, biosensing \cite{homola2003present}, and ultrahigh resolution lasers \cite{noginov2009demonstration}.

\begin{figure}[htp]
    \includegraphics[scale=1, left]{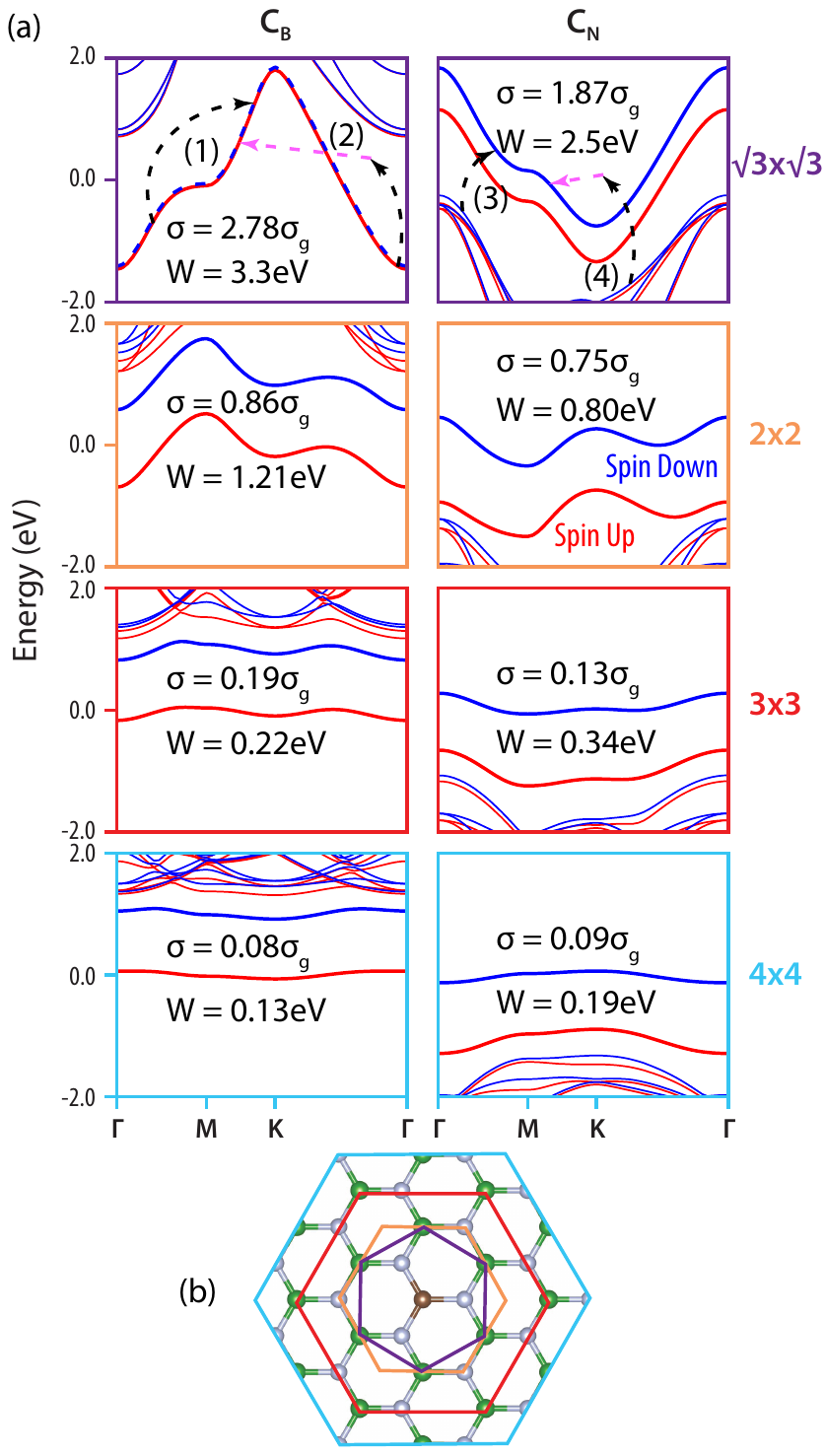}
    \caption{
    \textbf{Electronic bandstructure of carbon substitutional defect lattices in hexagonal boron nitride.} (a) Electronic bands for $C_B^{n\times n}$ (left column) and $C_N^{n\times n}$ (right column) structures. Energy is measured from the Fermi level. For the six lattices that are insulating and ferromagnetic, we shift the Fermi energy to half filling of the defect band closest to the hBN midgap. We report the bandwidth, $W$, and conductivity, $\sigma$, for each structure ($\sigma_g$ is graphene's conductivity at 0.5 eV doping). The labels (1) and (3) denote intraband and interband Landau damping processes, respectively. The labels (2) and (4) denote interband and interband phonon-assisted damping processes, respectively. Black dashed lines denote plasmons and pink dashed lines denote phonons. (b) The Wigner-Seitz cells for the 4 structures we  considered, within the largest, $4\times 4$ structure.  The brown circle denotes the C defect at the center of the cell. White circles denote N atoms for the B substitutional lattices and B atoms for the N substitutional lattices. Green circles denote B atoms for the B substitutional lattices and N atoms for the N substitutional lattices. The color scheme is the same as in (a) with purple, orange, red, and light blue corresponding to $\sqrt{3}\times\sqrt{3}$, $2\times 2$, $3\times 3$, and $4\times 4$, respectively.}  \label{fig: Bands}
\end{figure}

Our study is motivated by the goal of eliminating, or significantly reducing, the ubiquitous losses intrinsic to plasmonic materials \cite{khurgin2015deal, gjerding2017band}. 
In particular, we focus on 2D materials whose metallic character is \emph{induced} by the presence of defects 
(an example of case (iii) above). 
To avoid loss channels such as interband transitions, 
we restrict our search to a host material with a large band gap, namely hexagonal boron nitride (hBN) (6 eV bandgap, see \cite{cassabois2016hexagonal}), and defects that produce moderately flat bands near the middle of the band gap \footnote{A truly flat band does not support plasmonic excitations and a highly dispersive band increases intraband losses.}. For the defects, we choose carbon atoms, since their atomic size, similar to that of B and N,  minimizes defect-induced lattice strain. 
The substitutional defect structures we consider are denoted as $C_X^{n\times n}$, indicating an $n\times n$ supercell of the primitive unit cell of hBN, in which one atom labeled $X$ is replaced by a C atom (an example of which is shown in \cref{fig: Bands} (b)). 
We study supercells with $n = \sqrt{3}, 2, 3, 4$
and $X$=B or N, that is, a total of eight defect-containing structures. Typically, larger supercells can host flatter bands,  which are more conducive to low-loss plasmonics. 

The eight structures and corresponding electronic band structures are shown in \cref{fig: Bands}.  
Our DFT calculations indicate that the relaxed structures with N substitutional atoms are perfectly planar whereas those with B are slightly buckled;
see Supplementary Information (SI). While this is a minor structural difference, we show later that it has significant impact on the magnitude of the electron-phonon interaction. All of the structures, other than the two $C^{\sqrt{3}\times\sqrt{3}}_X$ ones, are fully spin polarized, meaning that they have to be doped in order to support plasmonic excitations. As previously predicted~\cite{weston2018native, liu2022spin}, we find that the structures with larger supercells are ferromagnetic with spin gaps on the order of $1$ eV. To verify their ferromangetic order, we calculate the ground state of a $2\times 2$ supercell of $C_{B}^{2\times 2}$ and find that the ferromagnetic state is preferred, with magnetization $4\mu_B$.
The structures we report have bandwidths ranging from 3.3 eV ($C_B^{\sqrt{3}\times\sqrt{3}}$) to 0.13 eV($C_B^{4\times4}$). This decrease in the bandwidth as lattice size is increased is accompanied by a commensurate decrease in the Fermi velocity and the onset of ferromagnetism (lifting of spin degeneracy), both of which contribute to a lowering of the Drude conductivity as seen in \cref{fig: Bands} (a). 
This leads to lower frequency plasmons for the larger defect supercells. 

Limiting plasmonic losses has been an active area of research for several decades \cite{giuliani1983charge}. 
In all proposed candidates for low-loss plasmonics, 
losses are mitigated by suppressing the phase space for direct and indirect (phonon-assisted) transitions into the electron-hole continuum. 
In the case of surface plasmons, for instance, one early proposal for avoiding losses was by engineering semiconducting superlattices that energetically separate the surface plasmonic band from the electron-hole continuum \cite{giuliani1983charge}.  In the case of our structures that include substitutional defects, a similar energetic separation should exist, as 
the structures with larger periodicity host flat bands that are well separated from the hBN valence and conduction bands. 
Accordingly, we calculate the 
transverse magnetic (TM) polarized plasmonic dispersions and associated losses in the proposed structures. 
For the $C_{B}^{\sqrt{3}\times\sqrt{3}}$ and $C_{N}^{\sqrt{3}\times\sqrt{3}}$ structures, 
which are metallic at charge neutrality, we calculate the plasmonic properties without imposing any changes in band occupation. 
For the structures with larger supercells, we move the Fermi level to half filling of the defect band closest to the hBN midgap and calculate plasmonic properties within the rigid band approximation \cite{lee2012validity}, which neglects changes to the band structure due to doping (we analyze the validity of this approximation below). Experimentally, this tuning of the Fermi level could be implemented by gating or chemical doping. 

To obtain the plasmonic dispersion for each case 
we evaluate the nonlocal, frequency dependent conductivity. Energies and wavefunctions used to calculate the conductivity are obtained from an {\em ab initio} tight binding model, derived from Wannier interpolation of the DFT band structures~\cite{sundararaman2020plasmonics}.
Specifically, we calculate the plasmon dispersion through the poles of the inverse dielectric function, $\varepsilon^{-1}(q, \omega)$,
which  at finite temperature is given by~\cite{mahan2013many}:
\begin{equation}
\frac{1}{\varepsilon(q, i\omega_n)} = 1 - \frac{V_q}{\Omega}\int_{0}^{\beta}  e^{i\omega_n \tau}
\braket{{\cal T}\rho(q, \tau)\rho(-q, 0)}d\tau, 
\label{equation: epsilon}
\end{equation}
where $\Omega$ is the unit cell area, ${\cal T}$ is the imaginary time ordering operator, $q$ is the wavevector and $V_q=\frac{e^2}{2\epsilon_0 q}$ is the Coulomb interaction in 2D (with $\epsilon_0$ the vacuum permittivity), $\beta$ is the inverse temperature, and $\rho(q, \tau)$ is the density operator in the Heisenberg representation. The above expression yields the inverse dielectric function at a bosonic Matsubara frequency, $\omega_n$. The retarded inverse dielectric function is then calculated by analytically continuing $i\omega_n \rightarrow \omega+i\delta$.

\begin{figure*}[htp]
\centering
    \includegraphics[scale=0.85]{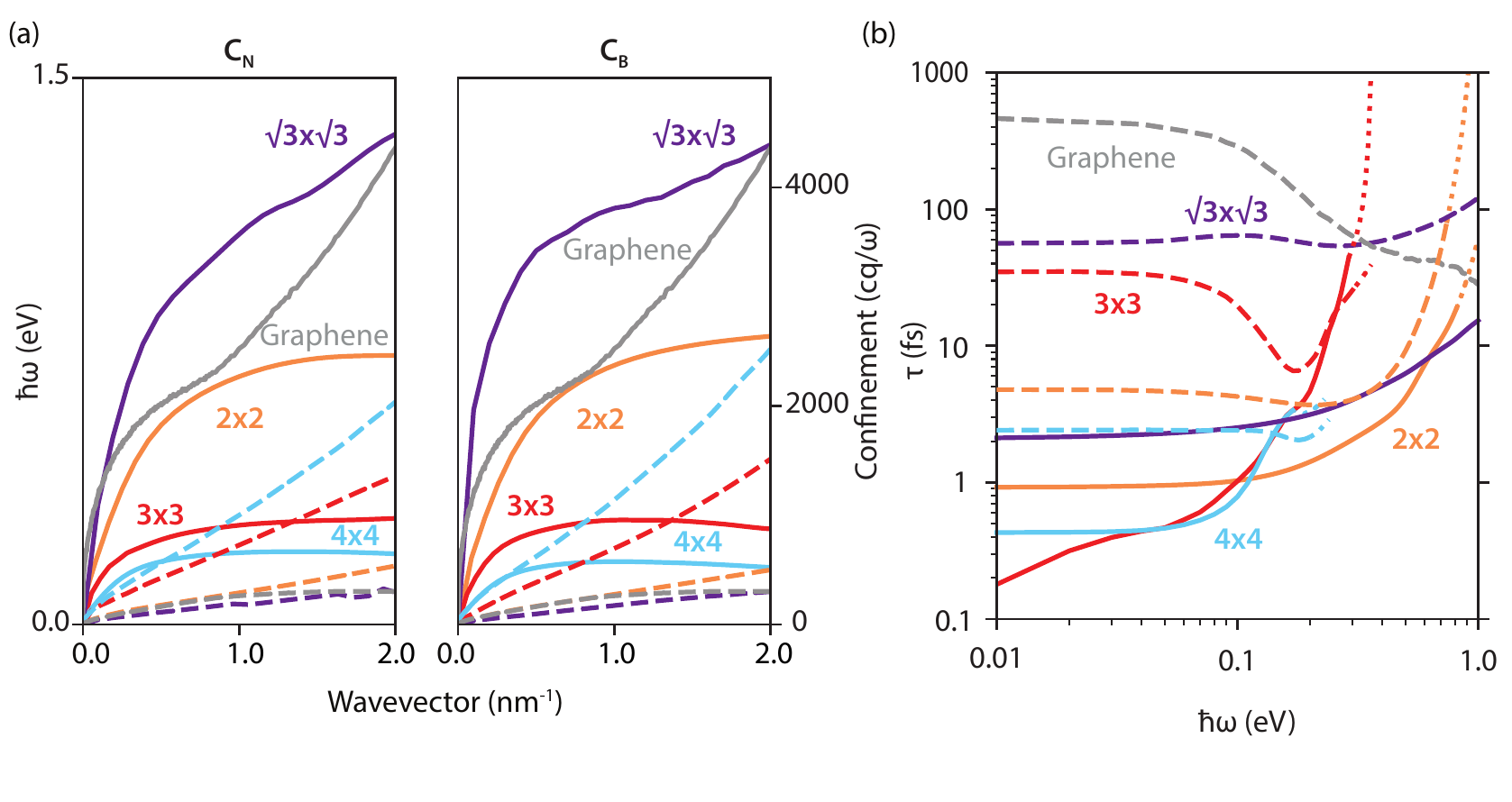}
    \caption{\textbf{Plasmonic dispersions, confinements and losses of carbon substitutional lattices in hexagonal boron nitride.} (a) Plasmonic dispersions (solid) and confinements (dashed) lines compared to graphene. Wavevectors were chosen to lie on the $\Gamma-M$ direction. (b) Decay times for $C_{B}^{n\times n}$ ($C_{N}^{n\times n}$) in solid (dashed) lines and graphene at 0.5 eV doping (dashed grey line). The point at which a solid (dashed) line turns into a dotted line indicates the highest plasmon frequency. The termination points of the dotted lines indicates the highest theoretical plasmon frequency (see discussion surrounding \cref{Equation: Enhancement}).}
    \label{fig: Plasmons}
\end{figure*}

Electron-electron interactions are included through the Random Phase Approximation (RPA) \cite{mahan2013many}, which gives the standard result (neglecting local-field effects) \cite{agarwal2014long} \footnote{In the present work, we disregarded local field effects~\cite{adler1962quantum, wiser1963dielectric}, which are likely substantial only when $|q| = |q+G|$, 
where $G$ is a reciprocal lattice vector. }: 
\begin{equation}
    \varepsilon(q, \omega) = 1-\frac{e^2}{2\varepsilon_0 q \Omega} {\cal F}(q,\omega)
    \label{equation:RPA_1}
\end{equation}
\begin{equation}
    {\cal F}(q,\omega) =
    \sum_{k} \frac{f(\epsilon^n_{k+q})-f(\epsilon^m_k)}{\epsilon^n_{k+q}-\epsilon^m_{k}-\hbar\omega-i\delta}|\braket{k+q, n| k, m}|^2, 
    \label{equation:RPA_2}
\end{equation}
where $\langle k+q, n|$, 
$| k, m\rangle$ denote
the cell periodic components 
of the Kohn-Sham eigenstates
with corresponding energy 
eigenvalues 
$\epsilon^n_{k+q}, \epsilon^m_k$. $f(\epsilon)$ is the Fermi occupation, with $\delta$ a small quantity to avoid singularities at the poles.
We note that in \cref{equation:RPA_2} the spin indices are subsumed into the band indices, $m,n$. We calculate $\varepsilon(q,\omega)$ 
from \cref{equation:RPA_1} and  \cref{equation:RPA_2} 
by first evaluating the imaginary part of this function and subsequently exploiting the Kramers-Kronig relations to compute the corresponding real part. We report plasmonic dispersions and confinements in \cref{fig: Plasmons} (a) as compared to the most well-established 2D plasmonic platform, graphene, at 0.5 eV doping from the Dirac point (0.5 eV doping puts the plasmons in the mid-IR). The plasmons in the proposed structures cover a frequency range from 0 to $\sim 1.3$ eV, with plasmons below 1 eV (see SI)  immune to interband and intraband losses. These plasmons have confinements in the plotted wavevector range of up to $\sim 8$  times that of graphene
(for the $C_{B}^{4\times 4}$ structure). We note also that the generically small plasmonic group velocities observed for most of the proposed structures is a consequence of interband screening, a nearly universal phenomenon in 2D~\cite{da2020universal}. 
This is not the case for the $C_B^{\sqrt{3}\times \sqrt{3}}$ and $C_N^{\sqrt{3}\times \sqrt{3}}$ structures, 
as in those two cases the plasmon approaches the interband continuum at large wavevectors $q$. 

To take into account the effect of phonons on the imaginary part of the dielectric function or, equivalently, the real part of the conductivity, we follow the prescription given in \cite{stauber2008effect, mahan2013many, allen2015electron} and evaluate the current-current response in the presence of electron-phonon interactions. We then translate this into a decay time in the Drude limit, using the relation: 
\begin{equation}
    \sigma(\omega) = \frac{e^2
    v_F^2 g(\epsilon_F)\tau(\omega)}{2\Omega(1-i\omega\tau(\omega))},
    \label{equation: Drude tau}
\end{equation}
 where $v_F$ is the Fermi velocity, $g(\epsilon_F)$ is the density of states per unit cell at the Fermi level and   $\tau(\omega)$ is the transport decay time (see SI). Through a Feynman diagrammatic expansion of the conductivity, the decay time can be rewritten in a style reminiscent of Fermi's golden rule \cite{allen1971electron, brown2016nonradiative}, giving an expression that is essentially a Fermi surface average of the 
 frequency-dependent carrier decay rate: 
\begin{multline}
        \tau^{-1}(\omega) = \frac{2\pi}{N_k N_{k'}\hbar^2\omega g(\epsilon_F)}\sum_{k, k' j \pm}
        |g_{k,k'}^{j}|^2\Big(N^{j, \mp}_{k-k'}f_k- \\ N_{k-k'}^{j, \pm}f_{k'} \pm f_kf_{k'} \Big) \delta(\epsilon_k+\hbar\omega\pm \hbar\omega^j_{k-k'}-\epsilon_{k'}) \Bigg(1-\frac{v_k \cdot v_{k'}}{|v_k||v_{k'}|}\Bigg), 
    \label{equation: decay rate}
\end{multline}
\normalsize
\noindent where 
$v_k$, $v_{k'}$ are the electronic velocities (corresponding to the intraband momentum matrix elements) at wavevectors $k$ and $k'$, respectively, $\omega_q^j$ are the phonon frequencies of branch $j$ at wavevector $q$,
with corresponding $N^j_q$ Bose occupation factors. For convenience, in the above equation we have defined the quantities $$N_q^{j, \pm} \equiv \frac{1}{2} +N_q^j \pm \frac{1}{2},$$ 
where the plus sign corresponds to phonon emission and the minus sign corresponds to phonon absorption. 
We sum over phonon bands, indexed by $j$, but we include only 
the defect-related  electronic band in the evaluation of the decay rate 
from 
\cref{equation: decay rate}, which is an exact expression in the frequency regime of interest (0-1 eV). 
We note that the combination of velocity factors in our expression, in the language of Feynman diagrams,
is a consequence of a vertex correction to the polarization bubble. In addition, we note that the combination of occupation factors in \cref{equation: decay rate} correctly takes into account reverse processes, as may be seen by the fact that the same combination of occupation factors appears in the time derivative of the phonon occupation factor, which vanishes at equilibrium due to detailed balance \cite{pines1962approach}. Lastly, we mention that
\cref{equation: decay rate} simplifies to the Fermi surface averaged electron-phonon decay rate in the limit $\hbar\omega \rightarrow 0$.

In \cref{fig: Plasmons} (b)
we present plasmonic decay times, $\tau(\omega)$. The $C_N^{\sqrt{3}\times \sqrt{3}}$ and  $C_N^{2\times 2}$lattices yield better plasmonic lifetimes than graphene at high frequencies, with $C_N^{\sqrt{3}\times \sqrt{3}}$ plasmons having lifetimes about four times those of graphene plasmons at $\sim 1$ eV. 
The fact that the decay times are lower for the B substitutional structures is due to enhancement of the electron-phonon interaction through the aformentioned buckling of the $C_B$ structures. We verified this by explicitly comparing the decay times for buckled and unbuckled (non-relaxed) $C_B$ structures (see SI). Though two $C_N$ lattices yield a small frequency range at which plasmonic quality is improved, it is clear that in general 
the proposed structures 
have shorter plasmonic decay times than graphene at most frequencies. This may be attributed to the fact that in the carrier decay rate, the density of electronic states effectively shows up twice in the numerator and only once in the denominator (SI). Thus, as a flat band hosts a tightly confined (in frequency) region with a high density of states, our observation of a small decay time at low frequencies is to be expected. 
In addition, the fact that flatter bands in general host lower frequency plasmons results in 
plasmons being pushed into the regime of high loss. 
This fact, however, 
offers insight that can lead
to a solution of the loss problem, posed as follows: 
If flatter bands yield a frequency range of low plasmonic loss that is above the actual plasmon frequency, how can the plasmon frequency
be pushed into this low-loss region? We investigate this question by determining, for each structure, 
their maximum theoretical
plasmon frequency, defined, for each $q$, as the highest possible frequency, $\omega$, for which the dielectric function 
$\varepsilon(q,\omega)$, defined 
in \cref{equation:RPA_1}, 
vanishes 
if we introduce a 
factor of $\alpha$, a positive constant, multiplying the quantity 
${\cal F}(q,\omega)$ 
defined in 
\cref{equation:RPA_2}.
Note that in the case of an infinite superlattice of 
structures stacked in the 
direction perpendicular to the planes ($z$-axis), this parameter 
takes the form
\begin{equation}
\alpha = \frac{\sinh(ql)}{\cosh(ql)-1}
\label{Equation: Enhancement}
\end{equation} 
where $l$ 
is the spacing of layers \cite{giuliani1983charge}. 
In \cref{fig: Plasmons} (b) 
we denote by dotted lines
the regime beyond the plasmon frequency and below the maximum theoretical plasmon frequency. 
Interestingly, for
the structures with larger periodicity, 
where interband polarization plays a dominant role, the plasmon frequency is 
not enhanced significantly in 
the stacked superlattices. 
Only for the 
case of $C_N^{2\times 2}$ and $C_B^{2\times 2}$ 
stacked superlattices is the plasmon dispersion enhanced significantly. 
To determine more accurately 
how much of an enhancement is possible, 
we calculate the plasmon dispersion of a $C_N^{2\times 2}$ superlattice with $l=1$ nm and determine that the plasmon frequency could reach $\approx 0.9$eV with a quality factor of 1100. 
Note that this result comes with a caveat: The reason that
a large value for $l$ is needed to obtain this effect, is that simply stacking the doped structures at their equilibrium interlayer distance 
$l \approx 0.3$ nm 
drastically changes the electronic structure. 
Thus, to actually achieve this effect experimentally, the doped layers  need to be separated  
by a spacer of low dielectric constant, 
such as pristine hBN. 
A systematic study of how the electronic structure would change in these many layered systems is beyond the scope of the present work. 


\begin{figure}[htp]
\centering
    \includegraphics[scale=1]{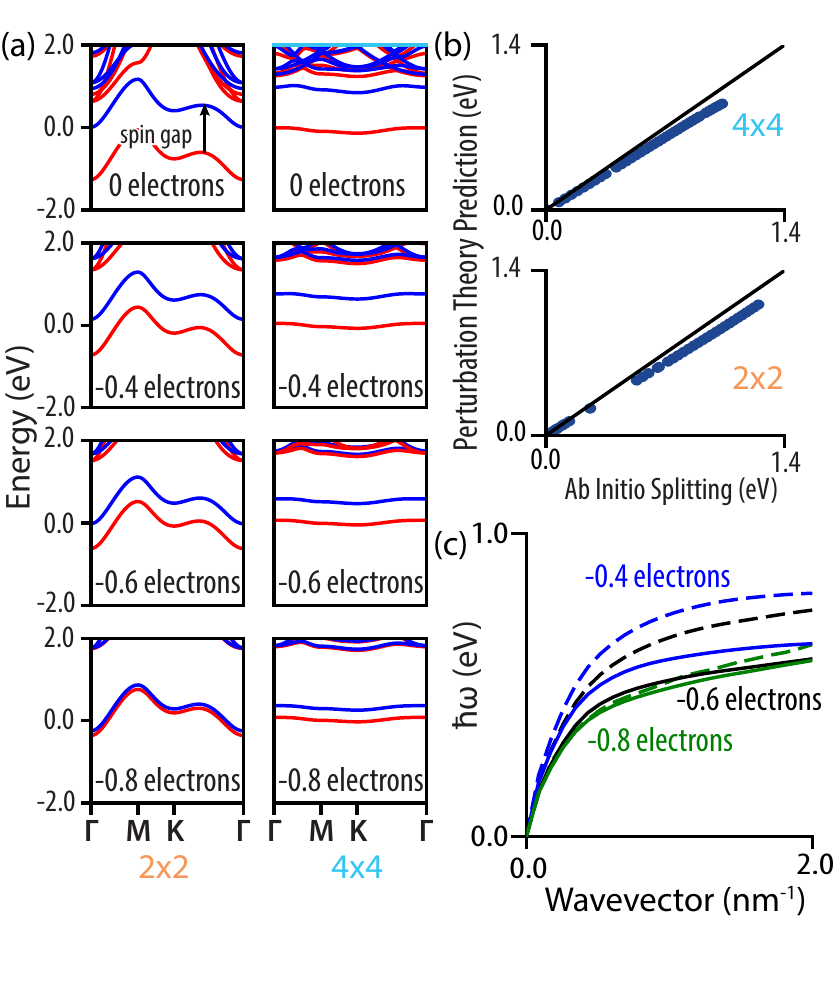}
    \caption{\textbf{Effect of doping on the spin-dependent bandstructures and plasmonic dispersion.} (a) Band structures near the Fermi level as a function of doping for $C^{2\times 2}_{B}$ (left column) and $C_{B}^{4\times4}$ (right column). (b) Accuracy of our spin splitting model given by \cref{equation: Perturbation theory}. Each blue dot represents a DFT calculation for which we plot our predicted spin splitting against the DFT result. The proximity of each point to the $y=x$ line (shown in black) indicates the validity of the perturbation theory model. (c) Plasmonic dispersion for $C_B^{2\times 2}$ at various values of doping calculated in the rigid band approximation (dashed lines) and from the explicit charged DFT calculations (solid lines)}
    \label{fig: Doping}
\end{figure}

We turn next to the validity of the rigid band approximation. Interestingly, we find that electron and hole doping in the structures 
we consider has the effect of 
tuning the spin-gap at the Fermi level (see \cref{fig: Doping} (a)), which is at odds with the bands being rigidly doped. We attribute this result to changes in the exchange potential with respect to doping. In particular, the exchange potential in DFT is a functional of the spin-resolved electron densities. From first order perturbation theory in the local density approximation (LDA), the
doping 
results in a spin gap given by
(see SI for details): 
\begin{equation}
\begin{split}
\Delta(\rho_{\uparrow}(r),
\rho_{\downarrow}(r)) = \Big(\frac{6}{\pi}\Big)^{1/3}\int (\rho_{\uparrow}^{1/3}(r)-\rho_{\downarrow}^{1/3}(r)) \times \\\frac{(|\psi_{\uparrow}(r)|^{2}+|\psi_{\downarrow}(r)|^{2})}{2} d^3r,
\end{split}
\label{equation: Perturbation theory}
\end{equation}
\normalsize
where $\rho_{\uparrow},\rho_{\downarrow}$
and $\psi_{\uparrow}, \psi_{\downarrow}$ 
are the spin up (down) densities and wavefunctions, respectively. 
We evaluate the reliability of this model by performing DFT calculations 
on the doped structures 
as shown in \cref{fig: Doping} (b). 
For each material, we use three 
functionals: PBE \cite{perdew1996generalized}, 
Slater LDA without correlation \cite{dirac1930note} 
and LDA with correlation \cite{perdew1981density}. 
For each value of doping, we calculate the spin splitting from DFT at the $\Gamma$ point (center of the Brillouin Zone in reciprocal space). 
We also use the self consistent density and the Kohn-Sham wavefunctions for the defect bands to calculate the predicted 
spin splitting value given by 
\cref{equation: Perturbation theory}. 
In \cref{fig: Doping}(b), we plot our prediction against the exact DFT values. The plots in \cref{fig: Doping}(b) 
 are obtained by using the PBE-GGA exchange correlation functional for the structures $C_{B}^{2\times 2}$ and $C_{B}^{4\times 4}$. As is evident, the fit of our model to the numerical DFT results is very accurate. 
We present similar plots for the other two exchange-correlation functions and for all other lattices in the SI. 
Note that the accurate fit we obtain in \cref{fig: Doping}(b) indicates that the effect of correlation and gradient terms in the Kohn-Sham Hamiltonian largely cancel out. Lastly, we note that a purely analytic estimate of the spin splitting 
may be carried out by decomposing the densities in terms of $1s$, $2s$, and $2p$ orbitals. We perform this calculation in the SI and obtain good agreement between this purely analytical calculation and our first principles results. 

In \cref{fig: Doping} (c)
we show plasmonic dispersions as a function of doping. 
We find that the explicitly doped plasmons differ qualitatively from those obtained through the rigid band approximation. 
At first this may seem like a consequence of the changing spin gap; however, we systematically investigated this issue 
and found that the major factor 
is actually a significant enhancement in interband wavefunction overlaps. Thus, the discrepancies 
in \cref{fig: Doping} (c) are primarily a consequence of the neutralizing charge background. 
To better understand how doping affects the plasmonic bands, 
other physically motivated scenarios
for doping, such as through lithium intercalation~\cite{profeta2012phonon}, 
need to be investigated.

In conclusion, 
we have introduced a set of candidate 2D materials with novel plasmonic properties, 
consisting of doped hBN through C substitution at either B or N sites. 
In particular, we predict these structures  to host plasmons with confinements up to eight times the maximum achievable in graphene, with decay times that can also surpass that of graphene, 
for frequency ranges exceeding $\sim 0.4$ eV. 
We have shown that stacking these 
structures could yield exceptionally high quality factors.  
We expect that imperfections in the periodicity of these materials to not 
have qualitative consequences 
as long as the density of defects is similar to the superlattices we investigated. However, this issue 
and the effect of impurities \cite{peres2008infrared} is a topic for future investigation. 
As most of the proposed structures  
have low Fermi velocities, 
it is possible that their electron-electron interactions necessitate an approach beyond RPA. 
Investigation of the validity of the random phase approximation in treating these flat band systems and their collective excitations warrants further investigation. 
Similar work which explored beyond RPA diagrams in the case of graphene~\cite{gangadharaiah2008charge}, has already been done but not applied to defect structures  of the type considered here. 
The effect of the electron-plasmon interaction on the carrier lifetimes should also be considered for a better assessment of the Drude decay time \cite{polini2008plasmons}. 
In addition, even in the absence of doping, exciton polaritons may exist in the proposed structures~\cite{henriques2022excitonic}, and needs to be further investigated. 
We note, in closing, that there has recently been a flurry of work in creating databases of 2D materials \cite{haastrup2018computational, gjerding2021recent}. 
While a ``blind'' enumeration of the plasmonic properties of all tabulated materials would be an overly demanding task, the results presented here suggest a simpler approach. Namely, we expect that filtering the available databases for materials with isolated flat bands at the Fermi level and high structural and thermal stability would be a first step in identifying the most promising candidates for low plasmonic losses. 

\paragraph{Computational Methods}
Density functional theory calculations were carried out with the use of the JDFTx package \cite{sundararaman2017jdftx} with norm-conserving pseudopotentials \cite{schlipf2015optimization} and Coulomb truncation for 2D materials \cite{sundararaman2013regularization}. The {\em ab initio} tight binding models used to calculate plasmonic properties were obtained by mapping onto a maximally localized Wannier basis \cite{souza2001maximally}. Ground state properties were calculated with a Fermi-level smearing of $10^{-5}$ Hartree, corresponding to a temperature of approximately 3 K. 
A higher smearing was used for the two metallic lattices due to the existence of a Fermi surface. 
Unless otherwise stated, the PBE exchange-correlation functional was used \cite{perdew1996generalized}. Comparisons between different functionals implemented in our modeling of doping dependent spin splitting additionally made use of pure Slater exchange \cite{dirac1930note} and the Perdew-Zunger local density approximation \cite{perdew1981density}. Ground state properties were found through either the self-consistent field method \cite{kresse1996efficient} or through electronic minimization \cite{freysoldt2009direct}. We show plots indicating convergence of the decay time results with respect to the phonon supercell size in SI. 

\paragraph{Acknowledgments}
This material is based upon work supported by the Air Force Office of Scientific Research under the award number
FA9550-21-1-0299, as well as in part by the U. S. Army Research Office through the Institute for Soldier Nanotechnologies at MIT, under Collaborative Agreement Number W911NF-18-2-0048. A.G. thanks the National Science Foundation Graduate Research Fellowship for financial support during the preparation of this paper. N.R. acknowledges the support of a Junior Fellowship from the Harvard Society of Fellows, as well as earlier support from a Computational Science Graduate Fellowship of the Department of Energy (DE-FG02-97ER25308), and a Dean’s Fellowship from the MIT School of Science. E.K. is supported in part by an
 Army Research Office grant under Cooperative Agreement Number W911NF-21-2-0147. The authors would also like to thank Ali Fahimniya, Cyprian Lewandowski, Thomas Christensen, Marinko Jablan, and Jennifer Coulter for useful discussions.
\onecolumngrid

\crefname{subequations}{Eqs.}{Eqs.} 
\Crefname{subequations}{Eqs.}{Eqs.}
\crefformat{subequations}{#2Eqs.~(#1)#3}
\Crefformat{subequations}{#2Eqs.~(#1)#3}
\crefname{page}{p.}{p.} 
\renewcommand{\thefigure}{S\arabic{figure}}
\renewcommand{\theequation}{S\arabic{equation}} 

\setcounter{equation}{0}
\setcounter{figure}{0}

\section{Supplementary Information for: Highly-confined and tunable plasmonics based on two-dimensional solid-state defect lattices}
\appendix

\section{Summary of Main Equations and Results}
Herein, we derive the main equations used in the main text and provide additional sanity checks, further motivating our results. First, we rigorously prove our RPA formulation for the dielectric function of our spinful systems. Next, we derive the expressions used to calculate the plasmonic dispersions, \cref{Plasmonic Dispersion}, and confinements. We next compare our decay time formula with an alternative expression , \cref{alternatetau}, reported in the literature \cite{kumar2022fermi}. Lastly, we lay out our formulation of doping dependent spin splitting, culminating in \cref{SpinSplitting} in the local density approximation. We also include a figure depicting the convergence of our decay time calculations with respect to phonon supercell size (\cref{Fig: Supercell Convergence}) and show how buckling of the B substitution lattices induces enhanced electron-phonon coupling in \cref{Fig: Buckled Decay}. 

\section{The Random Phase Approximation for Spinful Systems}

As the RPA is usually derived in a spin agnostic manner, we include a derivation below that retains the full spin structure. We remind the reader of the exact inverse dielectric function:

\begin{equation}
\small
    \epsilon^{-1}(q, i\omega_n) = 1-\frac{V_q}{\Omega} \sum_{k, k'}\int_0^\beta \braket{T(c_{k+q, \uparrow}^\dagger (\tau) c_{k, \uparrow}(\tau)+c_{k+q, \downarrow}^\dagger (\tau) c_{k, \downarrow}(\tau))(c_{k'-q, \uparrow}^\dagger(\tau') c_{k', \uparrow}(\tau') +c_{k'-q, \downarrow}^\dagger(\tau') c_{k', \downarrow}(\tau'))}e^{i\omega_n (\tau-\tau')}d(\tau-\tau'),
\end{equation}
where $V_q=\frac{e^2}{2\epsilon_0q}$ is the Coulomb interaction in 2D, q is the wavevector,  $\Omega$ is the unit cell area, $i\omega_n$ is a bosonic Matsubara frequency (we take $i\omega_n\rightarrow \omega+i\delta$ to get the retarded inverse dielectric function), and the $c_{k, (\downarrow,\uparrow)}$, $c^\dagger_{k, (\downarrow,\uparrow)}$ are spin-resolved Bloch state annihilation/creation operators. In the above, all operators are in the Heisenberg picture. We may write the above in terms of diagonal and off-diagonal (in spin indices) (interacting) density-density correlation functions: 

\begin{equation}
    \epsilon^{-1}(q, \omega) = 1+\frac{V_q}{\Omega}(\chi_{\uparrow \uparrow}(q, \omega)+\chi_{\downarrow \downarrow}(q, \omega)+\chi_{\downarrow \uparrow}(q, \omega)+\chi_{\uparrow \downarrow}(q, \omega))
\end{equation}

We write the RPA Dyson series for the diagonal up-up correlation function as (subsuming the area of the sample into the correlation function):

\begin{equation}
    \chi_{\uparrow \uparrow}(q, \omega) = \chi_{\uparrow \uparrow}^{0}(q, \omega) + V_q (\chi_{\uparrow \uparrow}^{0}(q, \omega))^2 + V_q^2 (\chi_{\uparrow \uparrow}^{0}(q, \omega))^2  (\chi_{\uparrow \uparrow}^{0}(q, \omega) + \chi_{\downarrow \downarrow}^{0}(q, \omega)) + ...
\end{equation}

At every order (after the zeroth order), we have two factors of the non-interacting up-up correlation function and one factor of $(\chi_{\uparrow \uparrow}^{0}(q, \omega) + \chi_{\downarrow \downarrow}^{0}(q, \omega))$. In the language of Feynman diagrams, this simply means that the end points of the RPA expansion have restricted spin indices whereas the intermediate vertices do not. The down-down correlation function may be similarly expanded as a Dyson series, and we do not reproduce it here. The up-down correlation function may be written as: 

\begin{equation}
    \chi_{\uparrow \downarrow}(q, \omega) = V_q \chi^0_{\uparrow \uparrow}(q, \omega)\chi^0_{\downarrow \downarrow}(q, \omega) + V_q^2 \chi^0_{\uparrow \uparrow}(q, \omega)\chi^0_{\downarrow \downarrow}(q, \omega)(\chi^0_{\uparrow \uparrow}(q, \omega) +\chi^0_{\downarrow \downarrow}(q, \omega)) + ...
\end{equation}

By the same logic as above, this Dyson series may be motivated by the fact that the endpoint vertices of the Feynman diagrammatic expansion have restricted spin indices (up and down, respectively). We may now readily sum up the Dyson series: 

\begin{equation}
    \chi_{\uparrow \uparrow}(q, \omega) = \chi^0_{\uparrow \uparrow}(q, \omega) + \frac{V_q \chi^0_{\uparrow \uparrow}(q, \omega)^2}{1-V_q(\chi^0_{\uparrow \uparrow}(q, \omega)+\chi^0_{\downarrow \downarrow}(q, \omega))}, \chi_{\uparrow \downarrow}(q, \omega) = \frac{V_q \chi^0_{\uparrow \uparrow} \chi^0_{\downarrow \downarrow}}{  1-V_q(\chi^0_{\uparrow \uparrow}(q, \omega)+\chi^0_{\downarrow \downarrow}(q, \omega))}
\end{equation}

Therefore, we have: 
\begin{equation}
      \epsilon^{-1}(q, \omega) =   1 + V_q\frac{\chi^0_{\uparrow \uparrow}(q, \omega)(1-V_q \chi^0_{\downarrow \downarrow}(q, \omega))+ \chi^0_{\downarrow \downarrow}(q, \omega)(1-V_q \chi^0_{\uparrow \uparrow}(q, \omega))+ 2\chi^0_{\uparrow \uparrow}(q, \omega))\chi_{\downarrow \downarrow}(q, \omega)) }{1-V_q(\chi_{\uparrow \uparrow}(q, \omega)+\chi_{\downarrow \downarrow}(q, \omega))}
\end{equation}

Which gives us for the dielectric function, 

\begin{equation}
    \epsilon(q, \omega) = 1-V_q(\chi^0_{\uparrow \uparrow}(q, \omega)+\chi^0_{\downarrow \downarrow}(q, \omega))
\end{equation}

This is the same result as derived in \cite{agarwal2014long}, although in that work the Dyson series is written in matrix form: 

\begin{equation}
    \chi^{-1}_{ij} = (\chi^0)^{-1}_{ij}-V_q
\end{equation}

\section{Plasmonic Dispersions}
We calculate TM polarized plasmons with mode profiles given as follows for the electric field (we choose the propagation direction of the plasmon to be in the x direction- other propagation directions follow trivially): 

\[
    E(x, y, z) = 
\begin{cases}
    e^{iqx -Qz} (e_x+\frac{iq}{Q}e_z),& \text{if } z > 0\\
    e^{iqx + Qz} (e_x-\frac{iq}{Q}e_z),& \text{if } z < 0 
\end{cases}
\]

To establish a relation between the frequency, $\omega$ and the wavevector, $q$, we calculate the current,  given by $J_x(q, \omega) = \sigma(q, \omega)E_x(q, \omega)$. This current carries a corresponding charge density $\rho(q, \omega) = \frac{q}{\omega}\sigma(q, \omega) E_x(q, \omega)$. Using Gauss's theorem, we obtain: 

\begin{equation}
    \frac{2iq}{Q} = \frac{q}{\omega\epsilon_0} \sigma(q, \omega) \rightarrow \frac{2i}{\sqrt{q^2-\omega^2/c^2}}=\frac{\sigma(q, \omega)}{\omega\epsilon_0}
    \label{Plasmonic Dispersion}
\end{equation}

\begin{figure*}
    \includegraphics[scale=1]{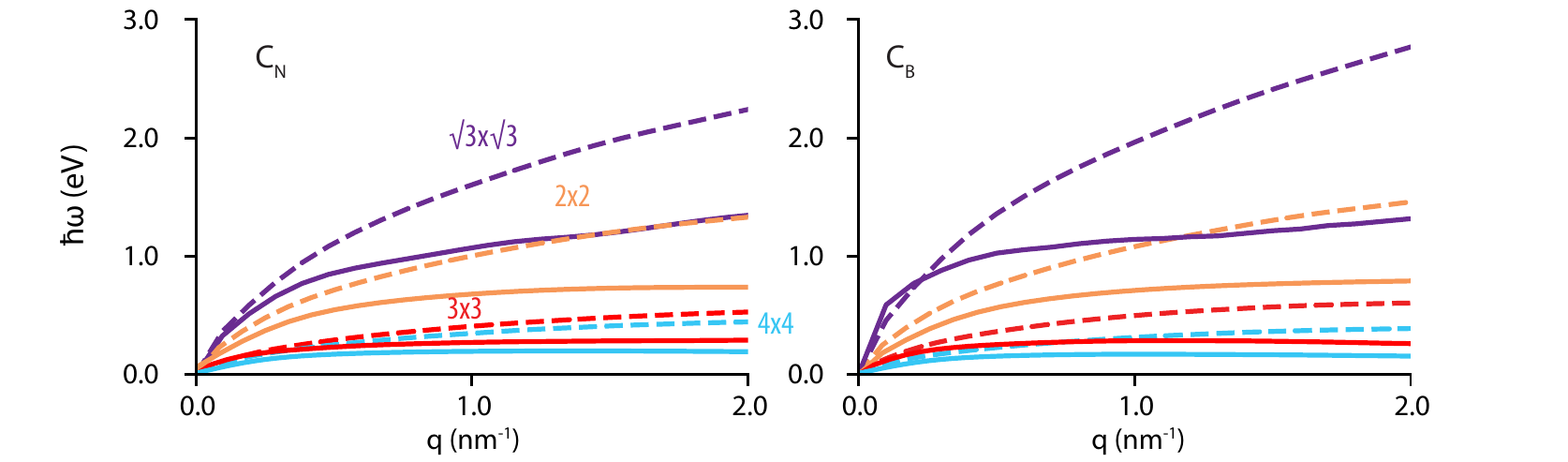}\caption{\textbf{Effective dielectric contrast.} Plasmonic dispersions derived purely from intraband transitions (dashed lines) and from both intraband and interband transitions (solid lines).}\label{IntervsIntra}
\end{figure*}

In principle, one must calculate the full non-local and frequency dependent conductivity- which is the way by which we obtain the reported plasmonic dispersions in the main text. To give an order of magnitude estimate of the plasmonic dispersion, however, we may calculate the Drude conductivity, given by: 

\begin{equation}
    \frac{ig_se^2}{N_k\omega\Omega}\sum_{k, n} \delta(\epsilon_{k}^n-\mu) |v^x_{n,k}|^2, 
\end{equation}
where $v_{n, k}^i$ is the velocity of band $n$ at wavevector $k$ in direction $i$, $\Omega$ is the unit cell area, $g_s$ is the spin degeneracy- which is 1 in the case of our defect lattices larger than $\sqrt{3}\times\sqrt{3}$ and 2 in the case of the two $\sqrt{3}\times\sqrt{3}$ lattices- and $N_k$ is the number of k points sampled in the Brillouin zone. We find that the local, Drude, conductivities of $C_B^{\sqrt{3}\times\sqrt{3}}$, $C_B^{2\times 2}$, $C_B^{3\times 3}$, $C_B^{4\times 4}$ are, respectively, $278\%$, $86\%$, $18.5\%$, $7.5\%$ that of graphene at $0.5$ eV doping, and the Drude conductivites for$C_N^{\sqrt{3}\times\sqrt{3}}$,  $C_N^{2\times 2}$, $C_N^{3\times 3}$, $C_N^{4\times 4}$ are, respectively, $187\%$, $75\%$, $13\%$, $8.7\%$ that of graphene at 0.5 eV doping. Physically, this trend reflects the fact that larger superlattices host progressively flatter bands- with lower Fermi velocities- at the chemical potential.
We note, also, that our lattices have a background effective dielectric contrast given by: 

\begin{equation}
    \epsilon(q) \approx 1+c\alpha_{BN}q
    \label{insulator epsilon}
\end{equation}
where $\alpha_{BN}$ is 12.69 in atomic units \cite{lonvcaric2018strong} and $c$ ranges from $1.6$ to $3.5$ depending on the particular superlattice. We confirm the value given for $\alpha_{BN}$ through a DFT calculation of pure hBN and obtain the range of values for $c$ by calculating how interband transitions alter the plasmonic dispersion in \cref{IntervsIntra}.

\section{Decay Times}\label{Decay Times}
\begin{figure*}
    \centering
    \includegraphics[scale=1]{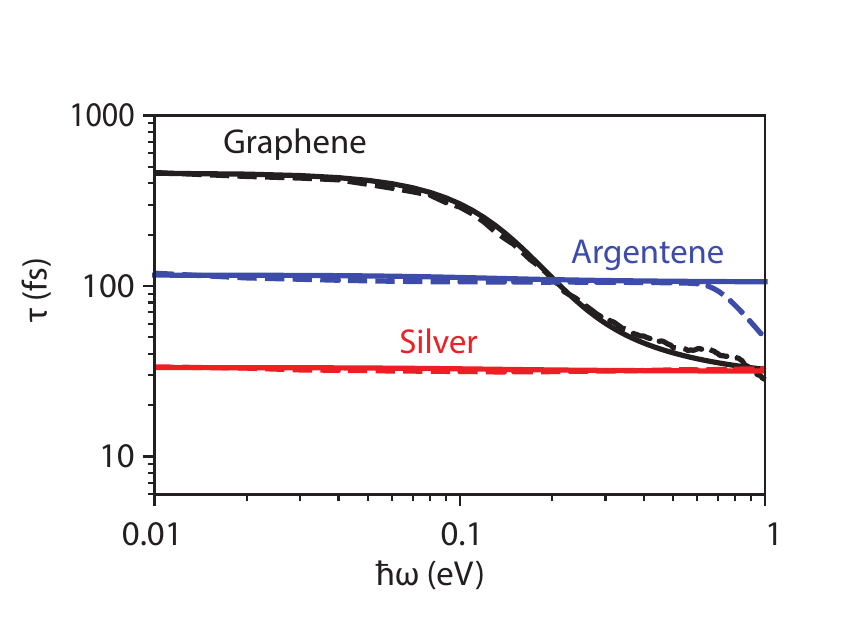}
    \caption{\textbf{Comparison of different decay times}. Dashed lines are obtained through the formula for $\tau(\omega)$ given in the main text, while the solid lines are given by the alternative decay time formula $\tau_{FS}(\omega)$ in \cref{alternatetau}. The discrepancy at high frequencies for Argentene is attributed to an interband transition that invalidates the Fermi surface approximation. }
    \label{taus}
\end{figure*}

\begin{figure*}
    \centering
    \includegraphics[scale=1]{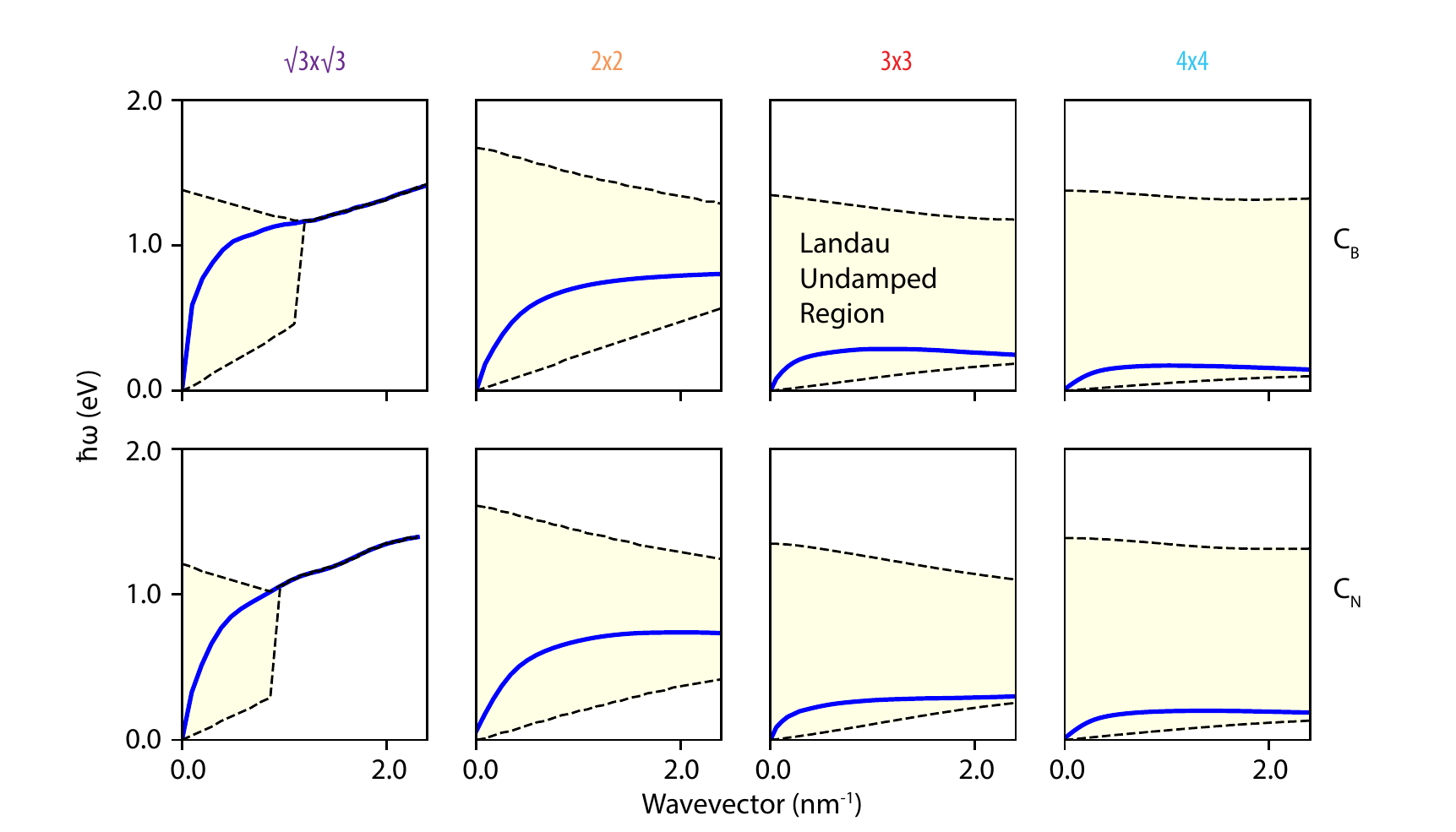}
    \caption{\textbf{Landau undamped plasmons.} Plasmonic dispersions (blue lines) sandwiched between the lower and upper bounds of the Landau damping region. In all cases, the lower (upper) bound corresponds to intraband (interband) electron-hole transitions.}
    \label{landau}
\end{figure*}

The lowest order of plasmon decay in our defect lattices is through the electron-phonon interaction. This is because all six plasmon dispersions are Landau undamped, as verified in \cref{landau}. 
In order to solidify the validity of the decay time formula presented in the main text, we compare decay times obtained through our formula with those obtained through: 
\begin{equation}
    \tau^{-1}_{FS}(\omega)=\frac{1}{\hbar g(\epsilon_F)} \frac{2\pi}{N_{k'}N_k}\sum_{k, k'} |g_{k, k'}^{n, m, \alpha}|^2\Big(1-\frac{v_k^n.v_{k'}^m}{|v_{k}^n||v_{k'}^m|}\Big)\delta(\epsilon_k^n-\epsilon_F)\delta(\epsilon_{k'}^m-\epsilon_F) \sum_{\pm}\frac{\pm b(\omega\pm \omega_q^\alpha)}{(e^{\pm\beta\hbar\omega_{q}^\alpha}-1)b(\omega)}
\label{alternatetau}\end{equation}
Where $\beta(\omega) = \frac{\beta\hbar\omega}{1-e^{-\beta\hbar\omega}}$, $g_{k, k'}^{n, m, \alpha}$ is the electron-phonon matrix element corresponding to electronic bands $n$ and $m$ at wavevectors $k$ and $k'$, respectively, and a phonon in branch $\alpha$ at wavevector $k'-k$ (of frequency $\omega_{k'-k}^\alpha$). $\epsilon_k^n$, $\epsilon_{k'}^m$ are the electronic energies. This formula has been used previously \cite{kumar2022fermi} to calculate the frequency dependent carrier relaxation time. We compare decay times obtained through our equation with those obtained through \cref{alternatetau} in \cref{taus} and obtain good agreement. The discrepancy for argentene at high frequencies is attributed to additional interband transitions that are not captured in \cref{alternatetau}.

A natural question to ask is how the above decay time is related to the plasmonic decay time as defined in \cite{wunsch2006dynamical}. To make this connection, we write the plasmonic decay rate as \cite{wunsch2006dynamical}: 

\begin{equation}
    \frac{\Re(\sigma(\omega)/\omega)}{\frac{\partial\Im(\sigma(\omega)/\omega}{\partial \omega}} = 1/2\tau(\omega)
 \end{equation}


\section{Doping Dependent Spin Splitting Model}\label{sec:Spin Splitting}

We start with the expression for the exchange energy in the local density approximation (LDA)

\begin{equation}
    E_x[\rho_{\uparrow}, \rho_{\downarrow}] = \frac{1}{2}\Big(E_x[2\rho_{\uparrow}]+ E_{x}[2\rho_{\downarrow}] \Big),
\end{equation}

where $\rho_{\uparrow}, \rho_{\downarrow}$ are the spin-resolved electronic densities. We write the total electronic density as $\rho$ in the following. In the absence of spin polarization, we have $\rho_{\uparrow} =\rho_{\downarrow} =\rho/2$, and we obtain the same exchange energy as when we neglect the existence of the two spin species. To determine how the Kohn-Sham eigenvalues change- as a function of doping- we must examine the exchange part of the Kohn-Sham potential. The matrix of exchange potentials is given by \cite{giustino2014materials}: 

\begin{equation}
    V_x^{\alpha\beta}  = \frac{\delta E_x}{\delta \rho_{\alpha \beta}} =\delta_{\alpha\beta} \frac{\delta E_x}{\delta \rho_{\alpha} },
\end{equation}

Where we define: 

\begin{equation}
    \rho_{\alpha \beta} = \sum_{i}\psi_{i, \alpha}^*\psi_{i, \beta}
\end{equation}

Where $i$ indexes all non-spin quantum numbers (such as band and crystal momentum) and $\alpha, \beta$ are spin indices (note also that, by definition, $\rho_{\uparrow \uparrow} = \rho_{\uparrow}$ and $\rho_{\downarrow \downarrow} = \rho_{\downarrow}$. By first order perturbation theory, the spin splitting is given by (in atomic units): 

\begin{equation}
   \Delta =  \int \Big(V^{\uparrow}(r) -V^{\downarrow}(r) \Big) \frac{|\psi_{\uparrow}(r)|^2+|\psi_{\downarrow}(r)|^2}{2} d^3r = \Big(\frac{3}{\pi} \Big)^{1/3}2^{1/3}\int (\rho_{\uparrow}^{1/3}(r)-\rho_{\downarrow}^{1/3}(r))\frac{|\psi_{\uparrow}(r)|^2+|\psi_{\downarrow}(r)|^2}{2} d^3r
\label{SpinSplitting}
\end{equation}

\begin{figure*}
    \includegraphics{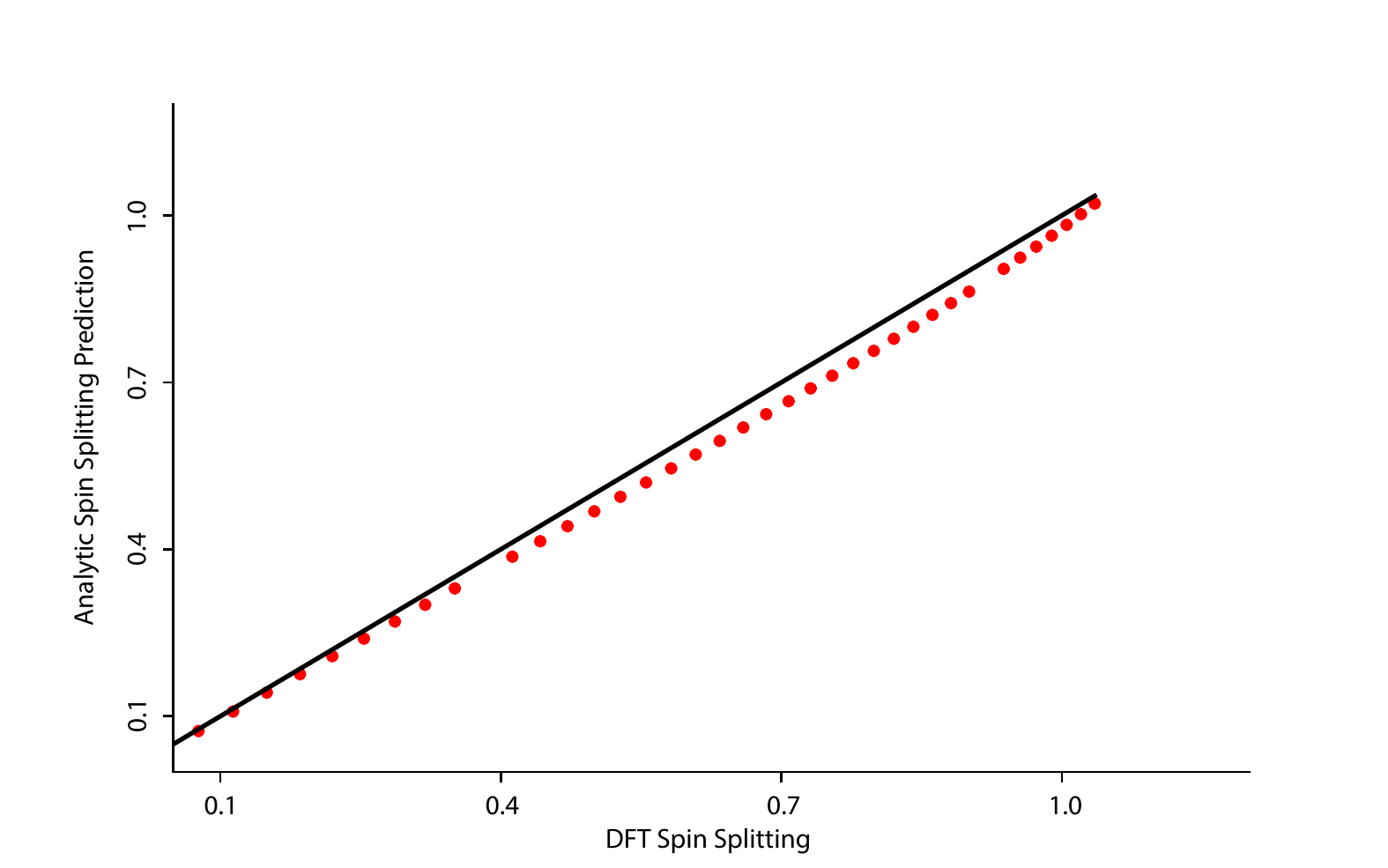}
    \caption{\textbf{Accuracy of analytic spin splitting model.} Analytic spin splitting model plotted against the result obtained from DFT through PBE exchange-correlation. Proximity to line at 45 degrees (black line) indicates accuracy of the model.}
    \label{Splitting-Analytic}
\end{figure*}

We first provide a fully analytical model for the spin splitting by using the fact that the electronic density interacting with the 2pz orbital at the Fermi level may be well described by a sum of densities contributed by the 1s, 2s  and the 2pz orbital itself. 

\begin{equation}
    \Big( \frac{6}{\pi} \Big)^{1/3} \int \Big[\Big( |\psi_{1s}(r)|^2 + |\psi_{2s}|^2 + (1-n)|\psi_{2pz}|^2\Big)^{1/3}-\Big( |\psi_{1s}(r)|^2 + |\psi_{2s}|^2 \Big)^{1/3} \Big] |\psi_{2pz}|^2 d^3r,
\end{equation}

where n is the absolute value of the excess charge. We numerically integrate this using the Hydrogenic wavefunctions for the 1s, 2s, and 2pz orbitals and obtain good agreement with our DFT results as shown in \cref{Splitting-Analytic}.

\begin{figure*}
    \centering
    \includegraphics[scale=1]{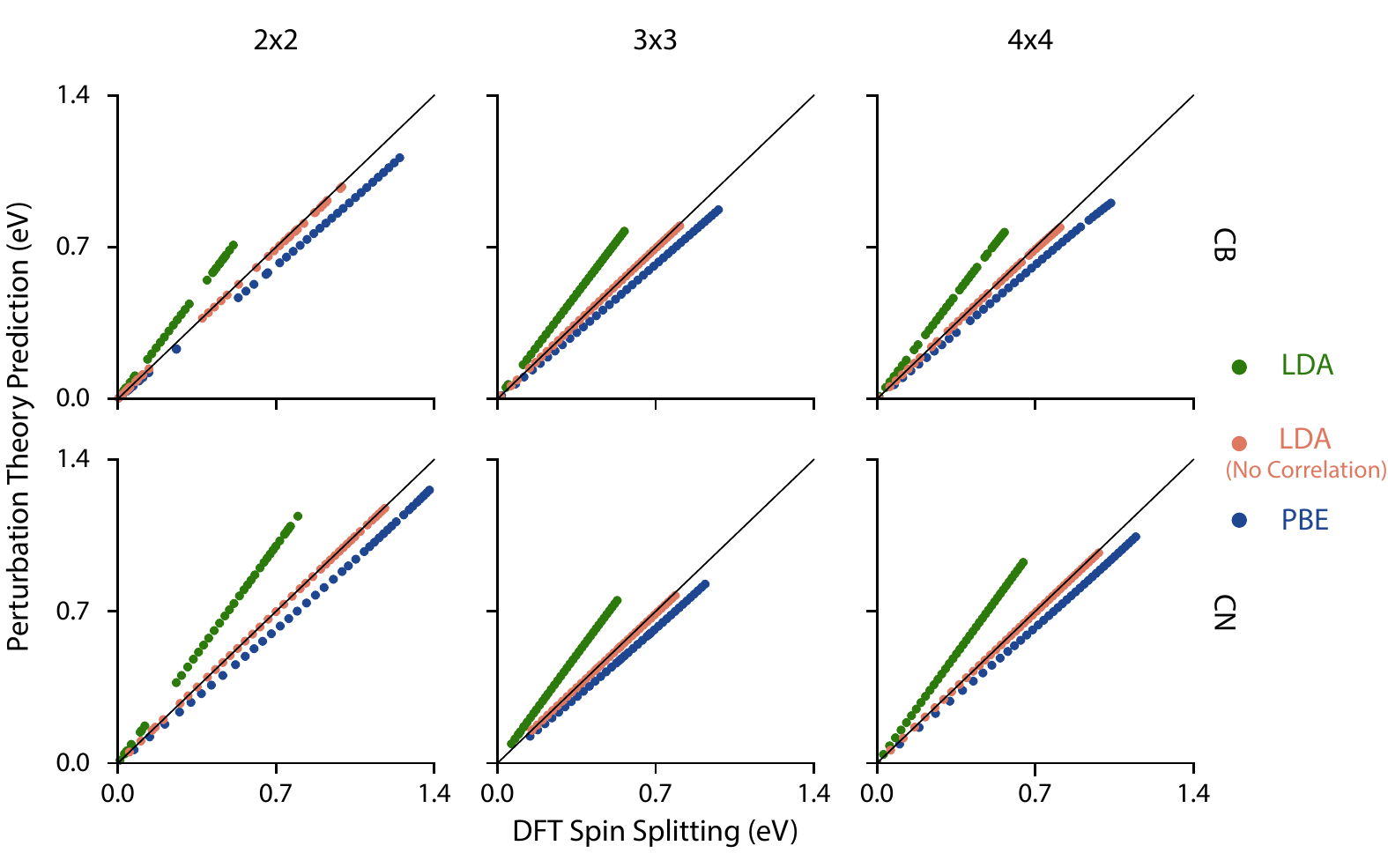}
    \caption{\textbf{Accuracy of first order perturbation theory in describing spin splitting.} Each point defines a unique DFT calculation. x axis defines the spin splitting as obtained through DFT, whereas the y axis is the result obtained through first order perturbation theory with the densities and wavefunctions obtained through DFT. The accuracy of the model is determined by how close a given point is to the solid line at 45 degrees.}
    \label{perturbation}
\end{figure*}

In \cref{perturbation}, we plot, for all six defect lattices, the result of \cref{SpinSplitting} against the DFT spin splitting result. We do this for three exchange-correlation functionals. Unsurpringly, LDA without correlation has the best fit with our model. This is motivated by the fact that our perturbation theory model uses the LDA exchange and neglects the effect of correlation. What is more surprising, however, is that the more accurate PBE exchange also yields good agreement with our model. 

\section{Hubbard Model For Spin Splitting}

We next show how the spin splitting may be predicted from the semianalytic evaluation of the Hubbard $U$ parameter. Explicitly, the Hubbard $U$ is given by: 

\begin{equation}
    U = \frac{e^2}{4\pi\epsilon_0}\int |\psi_{pz}(r_1)|^2\frac{1}{|r_1-r_2|} |\psi_{pz}(r_2)|^2 d^3r_1 d^3r_2
    \label{Hubbard}
\end{equation}

However, this model overestimates the actual $U$ parameter. This is because we have not taken into account screening. To take screening into account, we consider the case where we have full spin polarization, and, therefore, an insulator. In this case, the screened Coulomb interaction is given by (using \cref{insulator epsilon}): 

\begin{equation}
    W(\rho, \rho', z, z') = \frac{1}{(2\pi)^2}\int \frac{e^2e^{-iq|\rho-\rho'|\cos(\theta)} e^{-q|z-z'|}}{2\epsilon_0 q(1+\alpha q)} qdq d\theta
\end{equation}

We perform the integral over $\theta$, giving us: 

\begin{equation}
    \frac{e^2}{4\pi \epsilon_0}\int \frac{J_0(q|\rho-\rho'|)e^{-q|z-z'|}}{(1+\alpha q)}dq
\end{equation}

Therefore, the screened version of equation \ref{Hubbard} is: 

\begin{equation}
    \frac{e^2}{4\pi\epsilon_0}\int |\psi_{pz}(r_1)|^2 \frac{J_0(q|\rho-\rho'|)}{1+\alpha q} e^{-q|z-z'|}|\psi_{pz}(r_2)|^2 d^3r_1 d^3 r_2 dq 
\end{equation}

Evaluating for $\alpha \approx 20$ angstroms (a typical value for our defect lattices), gives us $U \approx 1.1-1.2$ eV, which is in good agreement with our ab-initio results of $\approx 1$ eV. 

\section{Buckling in Boron Substitution Lattices}

\begin{figure*}
    \centering
    \includegraphics[scale=1]{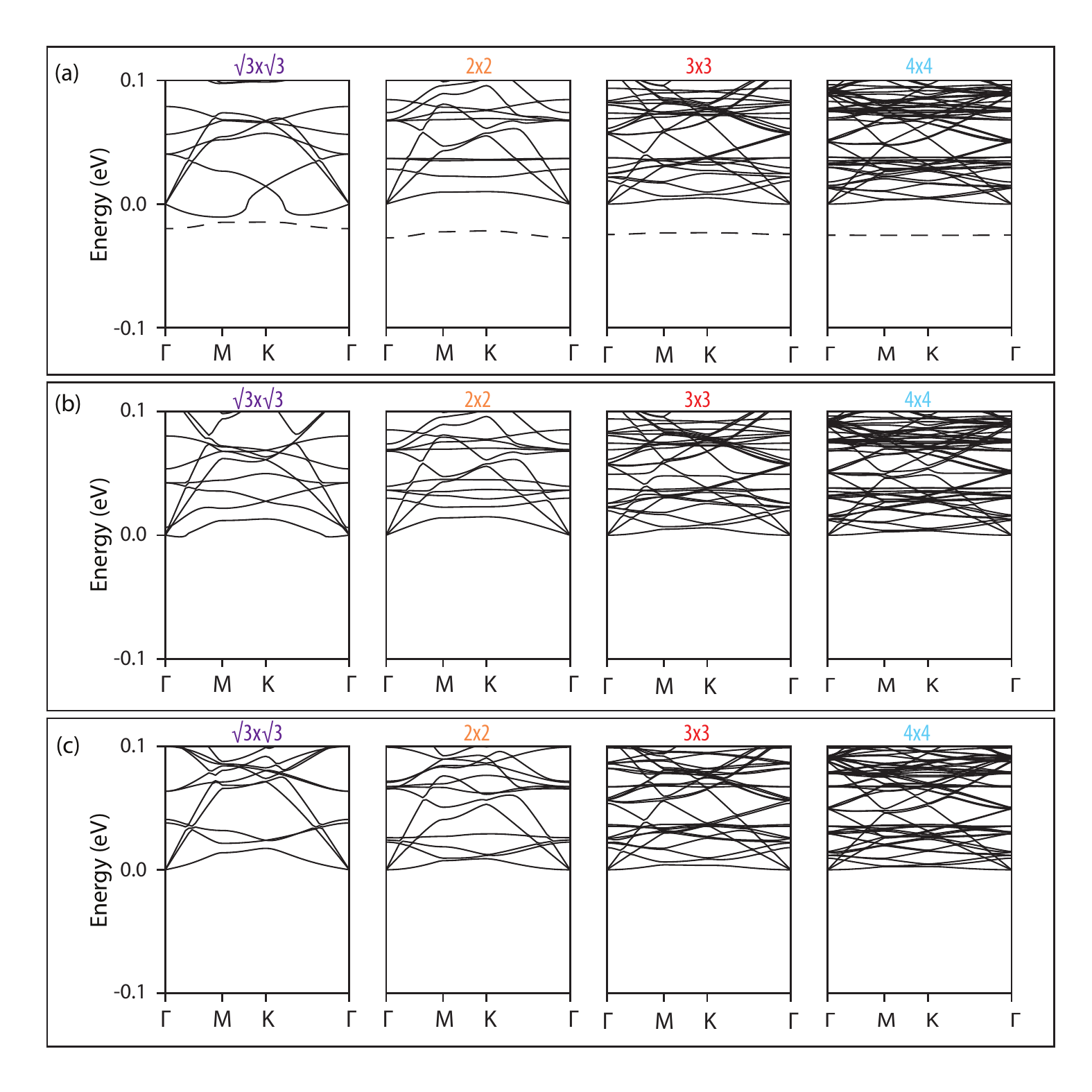}
    \caption{\textbf{Phonon dispersions and stability}. (a)  Phonons in planar systems of $C_B^{n\times n}$. The imaginary phonon modes shown as dashed lines indicate that a planar system is not stable. (b) Phonons of buckled $C_B^{n\times n}$ lattices. The lack of imaginary phonon frequencies indicates structural stability. Note that $C_B^{\sqrt{3}\times\sqrt{3}}$ has small imaginary phonon frequencies near $\Gamma$ on the order of 1meV. Since these are at incommensurate points in the Brillouin zone and their mangitudes get smaller as the phonon supercell is enlarged, they do not indicate a true instability. (c) Same as (a) but for $C_N^{n\times n}$ Lack of imaginary phonon frequencies indicates that planar $C_N^{n\times n}$ lattices are dynamically stable. }
    \label{buckle}
\end{figure*}

\begin{figure*}
    \centering
    \includegraphics[scale=1]{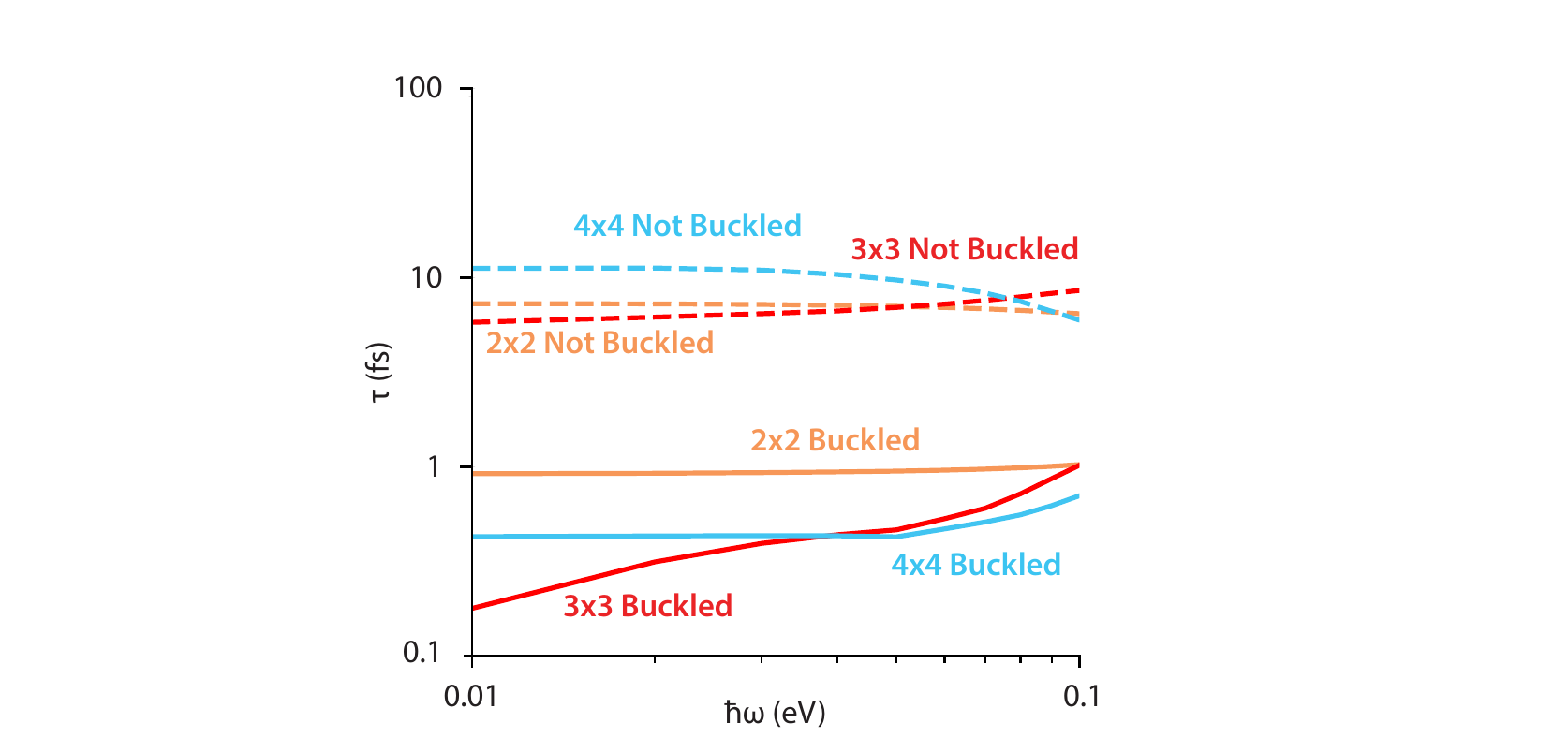}
    \caption{\textbf{Influence of buckling on electron-phonon coupling:} We calculate the decay time with the electron-phonon matrix elements of the relaxed buckled and unbuckled $C_B^{n\times n}$ systems. The generically higher decay times for the unbuckled systems indicate the enhancement of the electron-phonon coupling through buckling. }
    \label{Fig: Buckled Decay}
\end{figure*}

In \cref{buckle} we show the phonon dispersions of $C_{B}^{n\times n}$ for the case where lattice optimization is constrained to the x-y plane (top row) and for the case where the atoms are allowed to move out of plane (bottom row). The imaginary phonon frequencies in the top row subplots (shown as negative frequencies for convenience), indicate that all $C_{B}^{n\times n}$ lattices are buckled. We show in \cref{Fig: Buckled Decay} that the buckling leads to much lower decay times. 

\section{Plasmon Dispersion in Layered Systems}

We consider an infinite number of equally spaced planes of defect lattices (with lattice constant in the layered direction being $a$). To find the plasmon dispersion, it is sufficient to consider one interface, between regions 1 and 2. We take the ansatz $E_2 = e^{ika}E_1$, where $E_{1, 2}$ are the electric fields in regions 1 and 2, respectively. We take as the electric field in region 1: 

\begin{equation}
e^{iqx-i\omega t}(e^{-iQz}(Q, 0, q) + Ee^{iQz}(Q, 0, -q))    
\end{equation}

Continuity of the transverse part of the electric field at the interface requires:

\begin{equation}
    e^{ika}(1+E) = e^{-iQa}+Ee^{iQa} \rightarrow E = \frac{e^{ika}-e^{-iQa}}{e^{iQa}-e^{ika}}
\end{equation}

The H field in region 1 is given by: 

\begin{equation}
    H = -\frac{1}{\omega \mu}e^{iqx-i\omega t}\frac{\omega^2}{c^2}(e^{-iQz}+Ee^{iQz})e_y = -\omega \epsilon e^{iqx-i\omega t}(e^{-iQz}+Ee^{iQz})e_y
\end{equation}

Therefore, the change of the transverse part of the H field gives us: 


\begin{equation}
\epsilon (e^{ika}-e^{-iQa}+E(e^{ika}-e^{iQa})) = -\frac{iQ}{q^2}\Pi(q, \omega)(1+E)e^{ika}, 
\end{equation}

where $\Pi(q, \omega)$ is the polarizability \cite{jablan2009plasmonics}. Substituting for E and setting $Q=iq$, we get: 

\begin{equation}
    2\epsilon(e^{ika}-e^{qa})=\frac{1}{q}\Pi(q, \omega)\frac{e^{-qa}-e^{qa}}{e^{-qa}-e^{ika}}e^{ika} \rightarrow 1 = \frac{1}{2\epsilon q}\frac{e^{-qa}-e^{qa}}{e^{-qa}-e^{ika}-e^{-ika}+e^{qa}}\Pi(q,\omega)
\end{equation}

Therefore, we arrive at the final equation: 

\begin{equation}
1+\frac{\sinh(qa)}{\cosh(qa)-\cos(ka)}\frac{1}{2\epsilon q}\Pi(q, \omega) = 0
\end{equation}

\section{Convergence with Respect to Phonon Supercell Size}
\begin{figure*}
    \centering
    \includegraphics[scale=1]{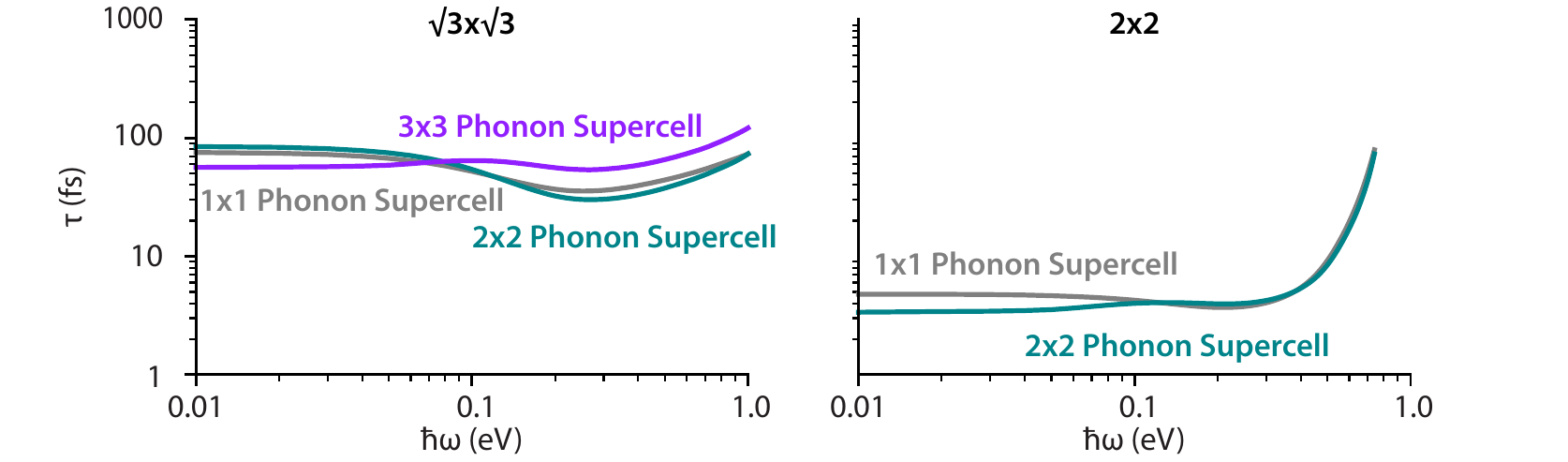}
    \caption{\textbf{Convergence of decay times with respect to phonon supercells:} We show the difference in decay times for two $C_N$ substitution lattices. As seen, the results do not change qualitatively with phonon supercell size.  }
    \label{Fig: Supercell Convergence}
\end{figure*}

In \cref{Fig: Supercell Convergence} we show the convergence with regards to supercell size for two of the defect lattices we studied. 

\bibliography{main.bib}

\begin{thebibliography}{53}
\expandafter\ifx\csname natexlab\endcsname\relax\def\natexlab#1{#1}\fi
\expandafter\ifx\csname bibnamefont\endcsname\relax
  \def\bibnamefont#1{#1}\fi
\expandafter\ifx\csname bibfnamefont\endcsname\relax
  \def\bibfnamefont#1{#1}\fi
\expandafter\ifx\csname citenamefont\endcsname\relax
  \def\citenamefont#1{#1}\fi
\expandafter\ifx\csname url\endcsname\relax
  \def\url#1{\texttt{#1}}\fi
\expandafter\ifx\csname urlprefix\endcsname\relax\def\urlprefix{URL }\fi
\providecommand{\bibinfo}[2]{#2}
\providecommand{\eprint}[2][]{\url{#2}}

\bibitem[{\citenamefont{Huang et~al.}(2022)\citenamefont{Huang, Grzeszczyk,
  Vaklinova, Watanabe, Taniguchi, Novoselov, and Koperski}}]{huang2022carbon}
\bibinfo{author}{\bibfnamefont{P.}~\bibnamefont{Huang}},
  \bibinfo{author}{\bibfnamefont{M.}~\bibnamefont{Grzeszczyk}},
  \bibinfo{author}{\bibfnamefont{K.}~\bibnamefont{Vaklinova}},
  \bibinfo{author}{\bibfnamefont{K.}~\bibnamefont{Watanabe}},
  \bibinfo{author}{\bibfnamefont{T.}~\bibnamefont{Taniguchi}},
  \bibinfo{author}{\bibfnamefont{K.}~\bibnamefont{Novoselov}},
  \bibnamefont{and} \bibinfo{author}{\bibfnamefont{M.}~\bibnamefont{Koperski}},
  \bibinfo{journal}{Physical Review B} \textbf{\bibinfo{volume}{106}},
  \bibinfo{pages}{014107} (\bibinfo{year}{2022}).

\bibitem[{\citenamefont{Lon{\v{c}}ari{\'c}
  et~al.}(2018)\citenamefont{Lon{\v{c}}ari{\'c}, Rukelj, Silkin, and
  Despoja}}]{lonvcaric2018strong}
\bibinfo{author}{\bibfnamefont{I.}~\bibnamefont{Lon{\v{c}}ari{\'c}}},
  \bibinfo{author}{\bibfnamefont{Z.}~\bibnamefont{Rukelj}},
  \bibinfo{author}{\bibfnamefont{V.~M.} \bibnamefont{Silkin}},
  \bibnamefont{and} \bibinfo{author}{\bibfnamefont{V.}~\bibnamefont{Despoja}},
  \bibinfo{journal}{npj 2D Materials and Applications}
  \textbf{\bibinfo{volume}{2}}, \bibinfo{pages}{1} (\bibinfo{year}{2018}).

\bibitem[{\citenamefont{Gjerding et~al.}(2021)\citenamefont{Gjerding,
  Taghizadeh, Rasmussen, Ali, Bertoldo, Deilmann, Kn{\o}sgaard, Kruse, Larsen,
  Manti et~al.}}]{gjerding2021recent}
\bibinfo{author}{\bibfnamefont{M.~N.} \bibnamefont{Gjerding}},
  \bibinfo{author}{\bibfnamefont{A.}~\bibnamefont{Taghizadeh}},
  \bibinfo{author}{\bibfnamefont{A.}~\bibnamefont{Rasmussen}},
  \bibinfo{author}{\bibfnamefont{S.}~\bibnamefont{Ali}},
  \bibinfo{author}{\bibfnamefont{F.}~\bibnamefont{Bertoldo}},
  \bibinfo{author}{\bibfnamefont{T.}~\bibnamefont{Deilmann}},
  \bibinfo{author}{\bibfnamefont{N.~R.} \bibnamefont{Kn{\o}sgaard}},
  \bibinfo{author}{\bibfnamefont{M.}~\bibnamefont{Kruse}},
  \bibinfo{author}{\bibfnamefont{A.~H.} \bibnamefont{Larsen}},
  \bibinfo{author}{\bibfnamefont{S.}~\bibnamefont{Manti}},
  \bibnamefont{et~al.}, \bibinfo{journal}{2D Materials}
  \textbf{\bibinfo{volume}{8}}, \bibinfo{pages}{044002} (\bibinfo{year}{2021}).

\bibitem[{\citenamefont{Lewandowski and
  Levitov}(2019)}]{lewandowski2019intrinsically}
\bibinfo{author}{\bibfnamefont{C.}~\bibnamefont{Lewandowski}} \bibnamefont{and}
  \bibinfo{author}{\bibfnamefont{L.}~\bibnamefont{Levitov}},
  \bibinfo{journal}{Proceedings of the National Academy of Sciences}
  \textbf{\bibinfo{volume}{116}}, \bibinfo{pages}{20869}
  (\bibinfo{year}{2019}).

\bibitem[{\citenamefont{Sundararaman et~al.}(2020)\citenamefont{Sundararaman,
  Christensen, Ping, Rivera, Joannopoulos, Solja{\v{c}}i{\'c}, and
  Narang}}]{sundararaman2020plasmonics}
\bibinfo{author}{\bibfnamefont{R.}~\bibnamefont{Sundararaman}},
  \bibinfo{author}{\bibfnamefont{T.}~\bibnamefont{Christensen}},
  \bibinfo{author}{\bibfnamefont{Y.}~\bibnamefont{Ping}},
  \bibinfo{author}{\bibfnamefont{N.}~\bibnamefont{Rivera}},
  \bibinfo{author}{\bibfnamefont{J.~D.} \bibnamefont{Joannopoulos}},
  \bibinfo{author}{\bibfnamefont{M.}~\bibnamefont{Solja{\v{c}}i{\'c}}},
  \bibnamefont{and} \bibinfo{author}{\bibfnamefont{P.}~\bibnamefont{Narang}},
  \bibinfo{journal}{Physical Review Materials} \textbf{\bibinfo{volume}{4}},
  \bibinfo{pages}{074011} (\bibinfo{year}{2020}).

\bibitem[{\citenamefont{Novoselov et~al.}(2005)\citenamefont{Novoselov, Jiang,
  Schedin, Booth, Khotkevich, Morozov, and Geim}}]{novoselov2005two}
\bibinfo{author}{\bibfnamefont{K.~S.} \bibnamefont{Novoselov}},
  \bibinfo{author}{\bibfnamefont{D.}~\bibnamefont{Jiang}},
  \bibinfo{author}{\bibfnamefont{F.}~\bibnamefont{Schedin}},
  \bibinfo{author}{\bibfnamefont{T.}~\bibnamefont{Booth}},
  \bibinfo{author}{\bibfnamefont{V.}~\bibnamefont{Khotkevich}},
  \bibinfo{author}{\bibfnamefont{S.}~\bibnamefont{Morozov}}, \bibnamefont{and}
  \bibinfo{author}{\bibfnamefont{A.~K.} \bibnamefont{Geim}},
  \bibinfo{journal}{Proceedings of the National Academy of Sciences}
  \textbf{\bibinfo{volume}{102}}, \bibinfo{pages}{10451}
  (\bibinfo{year}{2005}).

\bibitem[{\citenamefont{Boriskina et~al.}(2017)\citenamefont{Boriskina, Cooper,
  Zeng, Ni, Tong, Tsurimaki, Huang, Meroueh, Mahan, and
  Chen}}]{boriskina2017losses}
\bibinfo{author}{\bibfnamefont{S.~V.} \bibnamefont{Boriskina}},
  \bibinfo{author}{\bibfnamefont{T.~A.} \bibnamefont{Cooper}},
  \bibinfo{author}{\bibfnamefont{L.}~\bibnamefont{Zeng}},
  \bibinfo{author}{\bibfnamefont{G.}~\bibnamefont{Ni}},
  \bibinfo{author}{\bibfnamefont{J.~K.} \bibnamefont{Tong}},
  \bibinfo{author}{\bibfnamefont{Y.}~\bibnamefont{Tsurimaki}},
  \bibinfo{author}{\bibfnamefont{Y.}~\bibnamefont{Huang}},
  \bibinfo{author}{\bibfnamefont{L.}~\bibnamefont{Meroueh}},
  \bibinfo{author}{\bibfnamefont{G.}~\bibnamefont{Mahan}}, \bibnamefont{and}
  \bibinfo{author}{\bibfnamefont{G.}~\bibnamefont{Chen}},
  \bibinfo{journal}{Advances in Optics and Photonics}
  \textbf{\bibinfo{volume}{9}}, \bibinfo{pages}{775} (\bibinfo{year}{2017}).

\bibitem[{\citenamefont{Dai et~al.}(2019)\citenamefont{Dai, Fang, Rivera,
  Stehle, Jiang, Shen, Tay, Ciccarino, Ma, Rodan-Legrain
  et~al.}}]{dai2019phonon}
\bibinfo{author}{\bibfnamefont{S.}~\bibnamefont{Dai}},
  \bibinfo{author}{\bibfnamefont{W.}~\bibnamefont{Fang}},
  \bibinfo{author}{\bibfnamefont{N.}~\bibnamefont{Rivera}},
  \bibinfo{author}{\bibfnamefont{Y.}~\bibnamefont{Stehle}},
  \bibinfo{author}{\bibfnamefont{B.-Y.} \bibnamefont{Jiang}},
  \bibinfo{author}{\bibfnamefont{J.}~\bibnamefont{Shen}},
  \bibinfo{author}{\bibfnamefont{R.~Y.} \bibnamefont{Tay}},
  \bibinfo{author}{\bibfnamefont{C.~J.} \bibnamefont{Ciccarino}},
  \bibinfo{author}{\bibfnamefont{Q.}~\bibnamefont{Ma}},
  \bibinfo{author}{\bibfnamefont{D.}~\bibnamefont{Rodan-Legrain}},
  \bibnamefont{et~al.}, \bibinfo{journal}{Advanced materials}
  \textbf{\bibinfo{volume}{31}}, \bibinfo{pages}{1806603}
  (\bibinfo{year}{2019}).

\bibitem[{\citenamefont{Rivera et~al.}(2019)\citenamefont{Rivera, Christensen,
  and Narang}}]{rivera2019phonon}
\bibinfo{author}{\bibfnamefont{N.}~\bibnamefont{Rivera}},
  \bibinfo{author}{\bibfnamefont{T.}~\bibnamefont{Christensen}},
  \bibnamefont{and} \bibinfo{author}{\bibfnamefont{P.}~\bibnamefont{Narang}},
  \bibinfo{journal}{Nano Letters} \textbf{\bibinfo{volume}{19}},
  \bibinfo{pages}{2653} (\bibinfo{year}{2019}).

\bibitem[{\citenamefont{Novko et~al.}(2021)\citenamefont{Novko, Lyon, Mowbray,
  and Despoja}}]{novko2021ab}
\bibinfo{author}{\bibfnamefont{D.}~\bibnamefont{Novko}},
  \bibinfo{author}{\bibfnamefont{K.}~\bibnamefont{Lyon}},
  \bibinfo{author}{\bibfnamefont{D.~J.} \bibnamefont{Mowbray}},
  \bibnamefont{and} \bibinfo{author}{\bibfnamefont{V.}~\bibnamefont{Despoja}},
  \bibinfo{journal}{Physical Review B} \textbf{\bibinfo{volume}{104}},
  \bibinfo{pages}{115421} (\bibinfo{year}{2021}).

\bibitem[{\citenamefont{Khurgin}(2015{\natexlab{a}})}]{khurgin2015ultimate}
\bibinfo{author}{\bibfnamefont{J.~B.} \bibnamefont{Khurgin}},
  \bibinfo{journal}{Faraday discussions} \textbf{\bibinfo{volume}{178}},
  \bibinfo{pages}{109} (\bibinfo{year}{2015}{\natexlab{a}}).

\bibitem[{\citenamefont{Rivera et~al.}(2016)\citenamefont{Rivera, Kaminer,
  Zhen, Joannopoulos, and Solja{\v{c}}i{\'c}}}]{rivera2016shrinking}
\bibinfo{author}{\bibfnamefont{N.}~\bibnamefont{Rivera}},
  \bibinfo{author}{\bibfnamefont{I.}~\bibnamefont{Kaminer}},
  \bibinfo{author}{\bibfnamefont{B.}~\bibnamefont{Zhen}},
  \bibinfo{author}{\bibfnamefont{J.~D.} \bibnamefont{Joannopoulos}},
  \bibnamefont{and}
  \bibinfo{author}{\bibfnamefont{M.}~\bibnamefont{Solja{\v{c}}i{\'c}}},
  \bibinfo{journal}{Science} \textbf{\bibinfo{volume}{353}},
  \bibinfo{pages}{263} (\bibinfo{year}{2016}).

\bibitem[{\citenamefont{Rivera and Kaminer}(2020)}]{rivera2020light}
\bibinfo{author}{\bibfnamefont{N.}~\bibnamefont{Rivera}} \bibnamefont{and}
  \bibinfo{author}{\bibfnamefont{I.}~\bibnamefont{Kaminer}},
  \bibinfo{journal}{Nature Reviews Physics} \textbf{\bibinfo{volume}{2}},
  \bibinfo{pages}{538} (\bibinfo{year}{2020}).

\bibitem[{\citenamefont{Atwater and Polman}(2010)}]{atwater2010plasmonics}
\bibinfo{author}{\bibfnamefont{H.~A.} \bibnamefont{Atwater}} \bibnamefont{and}
  \bibinfo{author}{\bibfnamefont{A.}~\bibnamefont{Polman}},
  \bibinfo{journal}{Nature materials} \textbf{\bibinfo{volume}{9}},
  \bibinfo{pages}{205} (\bibinfo{year}{2010}).

\bibitem[{\citenamefont{Langer et~al.}(2019)\citenamefont{Langer, Jimenez~de
  Aberasturi, Aizpurua, Alvarez-Puebla, Augui{\'e}, Baumberg, Bazan, Bell,
  Boisen, Brolo et~al.}}]{langer2019present}
\bibinfo{author}{\bibfnamefont{J.}~\bibnamefont{Langer}},
  \bibinfo{author}{\bibfnamefont{D.}~\bibnamefont{Jimenez~de Aberasturi}},
  \bibinfo{author}{\bibfnamefont{J.}~\bibnamefont{Aizpurua}},
  \bibinfo{author}{\bibfnamefont{R.~A.} \bibnamefont{Alvarez-Puebla}},
  \bibinfo{author}{\bibfnamefont{B.}~\bibnamefont{Augui{\'e}}},
  \bibinfo{author}{\bibfnamefont{J.~J.} \bibnamefont{Baumberg}},
  \bibinfo{author}{\bibfnamefont{G.~C.} \bibnamefont{Bazan}},
  \bibinfo{author}{\bibfnamefont{S.~E.} \bibnamefont{Bell}},
  \bibinfo{author}{\bibfnamefont{A.}~\bibnamefont{Boisen}},
  \bibinfo{author}{\bibfnamefont{A.~G.} \bibnamefont{Brolo}},
  \bibnamefont{et~al.}, \bibinfo{journal}{ACS nano}
  \textbf{\bibinfo{volume}{14}}, \bibinfo{pages}{28} (\bibinfo{year}{2019}).

\bibitem[{\citenamefont{Homola}(2003)}]{homola2003present}
\bibinfo{author}{\bibfnamefont{J.}~\bibnamefont{Homola}},
  \bibinfo{journal}{Analytical and bioanalytical chemistry}
  \textbf{\bibinfo{volume}{377}}, \bibinfo{pages}{528} (\bibinfo{year}{2003}).

\bibitem[{\citenamefont{Noginov et~al.}(2009)\citenamefont{Noginov, Zhu,
  Belgrave, Bakker, Shalaev, Narimanov, Stout, Herz, Suteewong, and
  Wiesner}}]{noginov2009demonstration}
\bibinfo{author}{\bibfnamefont{M.}~\bibnamefont{Noginov}},
  \bibinfo{author}{\bibfnamefont{G.}~\bibnamefont{Zhu}},
  \bibinfo{author}{\bibfnamefont{A.}~\bibnamefont{Belgrave}},
  \bibinfo{author}{\bibfnamefont{R.}~\bibnamefont{Bakker}},
  \bibinfo{author}{\bibfnamefont{V.}~\bibnamefont{Shalaev}},
  \bibinfo{author}{\bibfnamefont{E.}~\bibnamefont{Narimanov}},
  \bibinfo{author}{\bibfnamefont{S.}~\bibnamefont{Stout}},
  \bibinfo{author}{\bibfnamefont{E.}~\bibnamefont{Herz}},
  \bibinfo{author}{\bibfnamefont{T.}~\bibnamefont{Suteewong}},
  \bibnamefont{and} \bibinfo{author}{\bibfnamefont{U.}~\bibnamefont{Wiesner}},
  \bibinfo{journal}{Nature} \textbf{\bibinfo{volume}{460}},
  \bibinfo{pages}{1110} (\bibinfo{year}{2009}).

\bibitem[{\citenamefont{Khurgin}(2015{\natexlab{b}})}]{khurgin2015deal}
\bibinfo{author}{\bibfnamefont{J.~B.} \bibnamefont{Khurgin}},
  \bibinfo{journal}{Nature nanotechnology} \textbf{\bibinfo{volume}{10}},
  \bibinfo{pages}{2} (\bibinfo{year}{2015}{\natexlab{b}}).

\bibitem[{\citenamefont{Gjerding et~al.}(2017)\citenamefont{Gjerding, Pandey,
  and Thygesen}}]{gjerding2017band}
\bibinfo{author}{\bibfnamefont{M.~N.} \bibnamefont{Gjerding}},
  \bibinfo{author}{\bibfnamefont{M.}~\bibnamefont{Pandey}}, \bibnamefont{and}
  \bibinfo{author}{\bibfnamefont{K.~S.} \bibnamefont{Thygesen}},
  \bibinfo{journal}{Nature communications} \textbf{\bibinfo{volume}{8}},
  \bibinfo{pages}{1} (\bibinfo{year}{2017}).

\bibitem[{\citenamefont{Cassabois et~al.}(2016)\citenamefont{Cassabois, Valvin,
  and Gil}}]{cassabois2016hexagonal}
\bibinfo{author}{\bibfnamefont{G.}~\bibnamefont{Cassabois}},
  \bibinfo{author}{\bibfnamefont{P.}~\bibnamefont{Valvin}}, \bibnamefont{and}
  \bibinfo{author}{\bibfnamefont{B.}~\bibnamefont{Gil}},
  \bibinfo{journal}{Nature photonics} \textbf{\bibinfo{volume}{10}},
  \bibinfo{pages}{262} (\bibinfo{year}{2016}).

\bibitem[{\citenamefont{Weston et~al.}(2018)\citenamefont{Weston,
  Wickramaratne, Mackoit, Alkauskas, and Van~de Walle}}]{weston2018native}
\bibinfo{author}{\bibfnamefont{L.}~\bibnamefont{Weston}},
  \bibinfo{author}{\bibfnamefont{D.}~\bibnamefont{Wickramaratne}},
  \bibinfo{author}{\bibfnamefont{M.}~\bibnamefont{Mackoit}},
  \bibinfo{author}{\bibfnamefont{A.}~\bibnamefont{Alkauskas}},
  \bibnamefont{and} \bibinfo{author}{\bibfnamefont{C.}~\bibnamefont{Van~de
  Walle}}, \bibinfo{journal}{Physical Review B} \textbf{\bibinfo{volume}{97}},
  \bibinfo{pages}{214104} (\bibinfo{year}{2018}).

\bibitem[{\citenamefont{Liu et~al.}(2022)\citenamefont{Liu, Guo, Yu, Meng, Li,
  Yang, Wang, Zeng, Xie, Wang et~al.}}]{liu2022spin}
\bibinfo{author}{\bibfnamefont{W.}~\bibnamefont{Liu}},
  \bibinfo{author}{\bibfnamefont{N.-J.} \bibnamefont{Guo}},
  \bibinfo{author}{\bibfnamefont{S.}~\bibnamefont{Yu}},
  \bibinfo{author}{\bibfnamefont{Y.}~\bibnamefont{Meng}},
  \bibinfo{author}{\bibfnamefont{Z.}~\bibnamefont{Li}},
  \bibinfo{author}{\bibfnamefont{Y.-Z.} \bibnamefont{Yang}},
  \bibinfo{author}{\bibfnamefont{Z.-A.} \bibnamefont{Wang}},
  \bibinfo{author}{\bibfnamefont{X.-D.} \bibnamefont{Zeng}},
  \bibinfo{author}{\bibfnamefont{L.-K.} \bibnamefont{Xie}},
  \bibinfo{author}{\bibfnamefont{J.-F.} \bibnamefont{Wang}},
  \bibnamefont{et~al.}, \bibinfo{journal}{Materials for Quantum Technology}
  (\bibinfo{year}{2022}).

\bibitem[{\citenamefont{Giuliani and Quinn}(1983)}]{giuliani1983charge}
\bibinfo{author}{\bibfnamefont{G.~F.} \bibnamefont{Giuliani}} \bibnamefont{and}
  \bibinfo{author}{\bibfnamefont{J.}~\bibnamefont{Quinn}},
  \bibinfo{journal}{Physical review letters} \textbf{\bibinfo{volume}{51}},
  \bibinfo{pages}{919} (\bibinfo{year}{1983}).

\bibitem[{\citenamefont{Lee and Mahanti}(2012)}]{lee2012validity}
\bibinfo{author}{\bibfnamefont{M.-S.} \bibnamefont{Lee}} \bibnamefont{and}
  \bibinfo{author}{\bibfnamefont{S.~D.} \bibnamefont{Mahanti}},
  \bibinfo{journal}{Physical Review B} \textbf{\bibinfo{volume}{85}},
  \bibinfo{pages}{165149} (\bibinfo{year}{2012}).

\bibitem[{\citenamefont{Mahan}(2013)}]{mahan2013many}
\bibinfo{author}{\bibfnamefont{G.~D.} \bibnamefont{Mahan}},
  \emph{\bibinfo{title}{Many-particle physics}} (\bibinfo{publisher}{Springer
  Science \& Business Media}, \bibinfo{year}{2013}).

\bibitem[{\citenamefont{Agarwal et~al.}(2014)\citenamefont{Agarwal, Polini,
  Vignale, and Flatt{\'e}}}]{agarwal2014long}
\bibinfo{author}{\bibfnamefont{A.}~\bibnamefont{Agarwal}},
  \bibinfo{author}{\bibfnamefont{M.}~\bibnamefont{Polini}},
  \bibinfo{author}{\bibfnamefont{G.}~\bibnamefont{Vignale}}, \bibnamefont{and}
  \bibinfo{author}{\bibfnamefont{M.~E.} \bibnamefont{Flatt{\'e}}},
  \bibinfo{journal}{Physical Review B} \textbf{\bibinfo{volume}{90}},
  \bibinfo{pages}{155409} (\bibinfo{year}{2014}).

\bibitem[{\citenamefont{da~Jornada et~al.}(2020)\citenamefont{da~Jornada, Xian,
  Rubio, and Louie}}]{da2020universal}
\bibinfo{author}{\bibfnamefont{F.~H.} \bibnamefont{da~Jornada}},
  \bibinfo{author}{\bibfnamefont{L.}~\bibnamefont{Xian}},
  \bibinfo{author}{\bibfnamefont{A.}~\bibnamefont{Rubio}}, \bibnamefont{and}
  \bibinfo{author}{\bibfnamefont{S.~G.} \bibnamefont{Louie}},
  \bibinfo{journal}{Nature communications} \textbf{\bibinfo{volume}{11}},
  \bibinfo{pages}{1013} (\bibinfo{year}{2020}).

\bibitem[{\citenamefont{Stauber and Peres}(2008)}]{stauber2008effect}
\bibinfo{author}{\bibfnamefont{T.}~\bibnamefont{Stauber}} \bibnamefont{and}
  \bibinfo{author}{\bibfnamefont{N.}~\bibnamefont{Peres}},
  \bibinfo{journal}{Journal of Physics: Condensed Matter}
  \textbf{\bibinfo{volume}{20}}, \bibinfo{pages}{055002}
  (\bibinfo{year}{2008}).

\bibitem[{\citenamefont{Allen}(2015)}]{allen2015electron}
\bibinfo{author}{\bibfnamefont{P.~B.} \bibnamefont{Allen}},
  \bibinfo{journal}{Physical Review B} \textbf{\bibinfo{volume}{92}},
  \bibinfo{pages}{054305} (\bibinfo{year}{2015}).

\bibitem[{\citenamefont{Allen}(1971)}]{allen1971electron}
\bibinfo{author}{\bibfnamefont{P.}~\bibnamefont{Allen}},
  \bibinfo{journal}{Physical Review B} \textbf{\bibinfo{volume}{3}},
  \bibinfo{pages}{305} (\bibinfo{year}{1971}).

\bibitem[{\citenamefont{Brown et~al.}(2016)\citenamefont{Brown, Sundararaman,
  Narang, Goddard~III, and Atwater}}]{brown2016nonradiative}
\bibinfo{author}{\bibfnamefont{A.~M.} \bibnamefont{Brown}},
  \bibinfo{author}{\bibfnamefont{R.}~\bibnamefont{Sundararaman}},
  \bibinfo{author}{\bibfnamefont{P.}~\bibnamefont{Narang}},
  \bibinfo{author}{\bibfnamefont{W.~A.} \bibnamefont{Goddard~III}},
  \bibnamefont{and} \bibinfo{author}{\bibfnamefont{H.~A.}
  \bibnamefont{Atwater}}, \bibinfo{journal}{ACS nano}
  \textbf{\bibinfo{volume}{10}}, \bibinfo{pages}{957} (\bibinfo{year}{2016}).

\bibitem[{\citenamefont{Pines and Schrieffer}(1962)}]{pines1962approach}
\bibinfo{author}{\bibfnamefont{D.}~\bibnamefont{Pines}} \bibnamefont{and}
  \bibinfo{author}{\bibfnamefont{J.~R.} \bibnamefont{Schrieffer}},
  \bibinfo{journal}{Physical Review} \textbf{\bibinfo{volume}{125}},
  \bibinfo{pages}{804} (\bibinfo{year}{1962}).

\bibitem[{\citenamefont{Perdew et~al.}(1996)\citenamefont{Perdew, Burke, and
  Ernzerhof}}]{perdew1996generalized}
\bibinfo{author}{\bibfnamefont{J.~P.} \bibnamefont{Perdew}},
  \bibinfo{author}{\bibfnamefont{K.}~\bibnamefont{Burke}}, \bibnamefont{and}
  \bibinfo{author}{\bibfnamefont{M.}~\bibnamefont{Ernzerhof}},
  \bibinfo{journal}{Physical review letters} \textbf{\bibinfo{volume}{77}},
  \bibinfo{pages}{3865} (\bibinfo{year}{1996}).

\bibitem[{\citenamefont{Dirac}(1930)}]{dirac1930note}
\bibinfo{author}{\bibfnamefont{P.~A.} \bibnamefont{Dirac}}, in
  \emph{\bibinfo{booktitle}{Mathematical proceedings of the Cambridge
  philosophical society}} (\bibinfo{organization}{Cambridge University Press},
  \bibinfo{year}{1930}), vol.~\bibinfo{volume}{26}, pp.
  \bibinfo{pages}{376--385}.

\bibitem[{\citenamefont{Perdew et~al.}(1981)\citenamefont{Perdew, McMullen, and
  Zunger}}]{perdew1981density}
\bibinfo{author}{\bibfnamefont{J.}~\bibnamefont{Perdew}},
  \bibinfo{author}{\bibfnamefont{E.}~\bibnamefont{McMullen}}, \bibnamefont{and}
  \bibinfo{author}{\bibfnamefont{A.}~\bibnamefont{Zunger}},
  \bibinfo{journal}{Physical Review A} \textbf{\bibinfo{volume}{23}},
  \bibinfo{pages}{2785} (\bibinfo{year}{1981}).

\bibitem[{\citenamefont{Profeta et~al.}(2012)\citenamefont{Profeta, Calandra,
  and Mauri}}]{profeta2012phonon}
\bibinfo{author}{\bibfnamefont{G.}~\bibnamefont{Profeta}},
  \bibinfo{author}{\bibfnamefont{M.}~\bibnamefont{Calandra}}, \bibnamefont{and}
  \bibinfo{author}{\bibfnamefont{F.}~\bibnamefont{Mauri}},
  \bibinfo{journal}{Nature physics} \textbf{\bibinfo{volume}{8}},
  \bibinfo{pages}{131} (\bibinfo{year}{2012}).

\bibitem[{\citenamefont{Peres et~al.}(2008)\citenamefont{Peres, Stauber, and
  Neto}}]{peres2008infrared}
\bibinfo{author}{\bibfnamefont{N.}~\bibnamefont{Peres}},
  \bibinfo{author}{\bibfnamefont{T.}~\bibnamefont{Stauber}}, \bibnamefont{and}
  \bibinfo{author}{\bibfnamefont{A.~C.} \bibnamefont{Neto}},
  \bibinfo{journal}{EPL (Europhysics Letters)} \textbf{\bibinfo{volume}{84}},
  \bibinfo{pages}{38002} (\bibinfo{year}{2008}).

\bibitem[{\citenamefont{Gangadharaiah et~al.}(2008)\citenamefont{Gangadharaiah,
  Farid, and Mishchenko}}]{gangadharaiah2008charge}
\bibinfo{author}{\bibfnamefont{S.}~\bibnamefont{Gangadharaiah}},
  \bibinfo{author}{\bibfnamefont{A.}~\bibnamefont{Farid}}, \bibnamefont{and}
  \bibinfo{author}{\bibfnamefont{E.}~\bibnamefont{Mishchenko}},
  \bibinfo{journal}{Physical review letters} \textbf{\bibinfo{volume}{100}},
  \bibinfo{pages}{166802} (\bibinfo{year}{2008}).

\bibitem[{\citenamefont{Polini et~al.}(2008)\citenamefont{Polini, Asgari,
  Borghi, Barlas, Pereg-Barnea, and MacDonald}}]{polini2008plasmons}
\bibinfo{author}{\bibfnamefont{M.}~\bibnamefont{Polini}},
  \bibinfo{author}{\bibfnamefont{R.}~\bibnamefont{Asgari}},
  \bibinfo{author}{\bibfnamefont{G.}~\bibnamefont{Borghi}},
  \bibinfo{author}{\bibfnamefont{Y.}~\bibnamefont{Barlas}},
  \bibinfo{author}{\bibfnamefont{T.}~\bibnamefont{Pereg-Barnea}},
  \bibnamefont{and}
  \bibinfo{author}{\bibfnamefont{A.}~\bibnamefont{MacDonald}},
  \bibinfo{journal}{Physical Review B} \textbf{\bibinfo{volume}{77}},
  \bibinfo{pages}{081411} (\bibinfo{year}{2008}).

\bibitem[{\citenamefont{Henriques et~al.}(2022)\citenamefont{Henriques, Amorim,
  Ribeiro, and Peres}}]{henriques2022excitonic}
\bibinfo{author}{\bibfnamefont{J.}~\bibnamefont{Henriques}},
  \bibinfo{author}{\bibfnamefont{B.}~\bibnamefont{Amorim}},
  \bibinfo{author}{\bibfnamefont{R.}~\bibnamefont{Ribeiro}}, \bibnamefont{and}
  \bibinfo{author}{\bibfnamefont{N.}~\bibnamefont{Peres}},
  \bibinfo{journal}{Physical Review B} \textbf{\bibinfo{volume}{105}},
  \bibinfo{pages}{115421} (\bibinfo{year}{2022}).

\bibitem[{\citenamefont{Haastrup et~al.}(2018)\citenamefont{Haastrup, Strange,
  Pandey, Deilmann, Schmidt, Hinsche, Gjerding, Torelli, Larsen, Riis-Jensen
  et~al.}}]{haastrup2018computational}
\bibinfo{author}{\bibfnamefont{S.}~\bibnamefont{Haastrup}},
  \bibinfo{author}{\bibfnamefont{M.}~\bibnamefont{Strange}},
  \bibinfo{author}{\bibfnamefont{M.}~\bibnamefont{Pandey}},
  \bibinfo{author}{\bibfnamefont{T.}~\bibnamefont{Deilmann}},
  \bibinfo{author}{\bibfnamefont{P.~S.} \bibnamefont{Schmidt}},
  \bibinfo{author}{\bibfnamefont{N.~F.} \bibnamefont{Hinsche}},
  \bibinfo{author}{\bibfnamefont{M.~N.} \bibnamefont{Gjerding}},
  \bibinfo{author}{\bibfnamefont{D.}~\bibnamefont{Torelli}},
  \bibinfo{author}{\bibfnamefont{P.~M.} \bibnamefont{Larsen}},
  \bibinfo{author}{\bibfnamefont{A.~C.} \bibnamefont{Riis-Jensen}},
  \bibnamefont{et~al.}, \bibinfo{journal}{2D Materials}
  \textbf{\bibinfo{volume}{5}}, \bibinfo{pages}{042002} (\bibinfo{year}{2018}).

\bibitem[{\citenamefont{Sundararaman et~al.}(2017)\citenamefont{Sundararaman,
  Letchworth-Weaver, Schwarz, Gunceler, Ozhabes, and
  Arias}}]{sundararaman2017jdftx}
\bibinfo{author}{\bibfnamefont{R.}~\bibnamefont{Sundararaman}},
  \bibinfo{author}{\bibfnamefont{K.}~\bibnamefont{Letchworth-Weaver}},
  \bibinfo{author}{\bibfnamefont{K.~A.} \bibnamefont{Schwarz}},
  \bibinfo{author}{\bibfnamefont{D.}~\bibnamefont{Gunceler}},
  \bibinfo{author}{\bibfnamefont{Y.}~\bibnamefont{Ozhabes}}, \bibnamefont{and}
  \bibinfo{author}{\bibfnamefont{T.}~\bibnamefont{Arias}},
  \bibinfo{journal}{SoftwareX} \textbf{\bibinfo{volume}{6}},
  \bibinfo{pages}{278} (\bibinfo{year}{2017}).

\bibitem[{\citenamefont{Schlipf and Gygi}(2015)}]{schlipf2015optimization}
\bibinfo{author}{\bibfnamefont{M.}~\bibnamefont{Schlipf}} \bibnamefont{and}
  \bibinfo{author}{\bibfnamefont{F.}~\bibnamefont{Gygi}},
  \bibinfo{journal}{Computer Physics Communications}
  \textbf{\bibinfo{volume}{196}}, \bibinfo{pages}{36} (\bibinfo{year}{2015}).

\bibitem[{\citenamefont{Sundararaman and
  Arias}(2013)}]{sundararaman2013regularization}
\bibinfo{author}{\bibfnamefont{R.}~\bibnamefont{Sundararaman}}
  \bibnamefont{and} \bibinfo{author}{\bibfnamefont{T.}~\bibnamefont{Arias}},
  \bibinfo{journal}{Physical Review B} \textbf{\bibinfo{volume}{87}},
  \bibinfo{pages}{165122} (\bibinfo{year}{2013}).

\bibitem[{\citenamefont{Souza et~al.}(2001)\citenamefont{Souza, Marzari, and
  Vanderbilt}}]{souza2001maximally}
\bibinfo{author}{\bibfnamefont{I.}~\bibnamefont{Souza}},
  \bibinfo{author}{\bibfnamefont{N.}~\bibnamefont{Marzari}}, \bibnamefont{and}
  \bibinfo{author}{\bibfnamefont{D.}~\bibnamefont{Vanderbilt}},
  \bibinfo{journal}{Physical Review B} \textbf{\bibinfo{volume}{65}},
  \bibinfo{pages}{035109} (\bibinfo{year}{2001}).

\bibitem[{\citenamefont{Kresse and
  Furthm{\"u}ller}(1996)}]{kresse1996efficient}
\bibinfo{author}{\bibfnamefont{G.}~\bibnamefont{Kresse}} \bibnamefont{and}
  \bibinfo{author}{\bibfnamefont{J.}~\bibnamefont{Furthm{\"u}ller}},
  \bibinfo{journal}{Physical review B} \textbf{\bibinfo{volume}{54}},
  \bibinfo{pages}{11169} (\bibinfo{year}{1996}).

\bibitem[{\citenamefont{Freysoldt et~al.}(2009)\citenamefont{Freysoldt, Boeck,
  and Neugebauer}}]{freysoldt2009direct}
\bibinfo{author}{\bibfnamefont{C.}~\bibnamefont{Freysoldt}},
  \bibinfo{author}{\bibfnamefont{S.}~\bibnamefont{Boeck}}, \bibnamefont{and}
  \bibinfo{author}{\bibfnamefont{J.}~\bibnamefont{Neugebauer}},
  \bibinfo{journal}{Physical Review B} \textbf{\bibinfo{volume}{79}},
  \bibinfo{pages}{241103} (\bibinfo{year}{2009}).

\bibitem[{\citenamefont{Kumar et~al.}(2022)\citenamefont{Kumar, Multunas, and
  Sundararaman}}]{kumar2022fermi}
\bibinfo{author}{\bibfnamefont{S.}~\bibnamefont{Kumar}},
  \bibinfo{author}{\bibfnamefont{C.}~\bibnamefont{Multunas}}, \bibnamefont{and}
  \bibinfo{author}{\bibfnamefont{R.}~\bibnamefont{Sundararaman}},
  \bibinfo{journal}{Physical Review Materials} \textbf{\bibinfo{volume}{6}},
  \bibinfo{pages}{125201} (\bibinfo{year}{2022}).

\bibitem[{\citenamefont{Wunsch et~al.}(2006)\citenamefont{Wunsch, Stauber,
  Sols, and Guinea}}]{wunsch2006dynamical}
\bibinfo{author}{\bibfnamefont{B.}~\bibnamefont{Wunsch}},
  \bibinfo{author}{\bibfnamefont{T.}~\bibnamefont{Stauber}},
  \bibinfo{author}{\bibfnamefont{F.}~\bibnamefont{Sols}}, \bibnamefont{and}
  \bibinfo{author}{\bibfnamefont{F.}~\bibnamefont{Guinea}},
  \bibinfo{journal}{New Journal of Physics} \textbf{\bibinfo{volume}{8}},
  \bibinfo{pages}{318} (\bibinfo{year}{2006}).

\bibitem[{\citenamefont{Giustino}(2014)}]{giustino2014materials}
\bibinfo{author}{\bibfnamefont{F.}~\bibnamefont{Giustino}},
  \emph{\bibinfo{title}{Materials modelling using density functional theory:
  properties and predictions}} (\bibinfo{publisher}{Oxford University Press},
  \bibinfo{year}{2014}).

\bibitem[{\citenamefont{Jablan et~al.}(2009)\citenamefont{Jablan, Buljan, and
  Solja{\v{c}}i{\'c}}}]{jablan2009plasmonics}
\bibinfo{author}{\bibfnamefont{M.}~\bibnamefont{Jablan}},
  \bibinfo{author}{\bibfnamefont{H.}~\bibnamefont{Buljan}}, \bibnamefont{and}
  \bibinfo{author}{\bibfnamefont{M.}~\bibnamefont{Solja{\v{c}}i{\'c}}},
  \bibinfo{journal}{Physical review B} \textbf{\bibinfo{volume}{80}},
  \bibinfo{pages}{245435} (\bibinfo{year}{2009}).

\bibitem[{\citenamefont{Adler}(1962)}]{adler1962quantum}
\bibinfo{author}{\bibfnamefont{S.~L.} \bibnamefont{Adler}},
  \bibinfo{journal}{Physical Review} \textbf{\bibinfo{volume}{126}},
  \bibinfo{pages}{413} (\bibinfo{year}{1962}).

\bibitem[{\citenamefont{Wiser}(1963)}]{wiser1963dielectric}
\bibinfo{author}{\bibfnamefont{N.}~\bibnamefont{Wiser}},
  \bibinfo{journal}{Physical Review} \textbf{\bibinfo{volume}{129}},
  \bibinfo{pages}{62} (\bibinfo{year}{1963}).

\end{thebibliography}
\end{document}